\documentclass{aa}
\usepackage{psfig,latexsym,amssymb}

\begin{document}
\title{Observing Mkn 421 with XMM--Newton: the EPIC--PN point of view}
\author{M.~Ravasio
\and G.~Tagliaferri
\and G.~Ghisellini 
\and F.~Tavecchio
}
\offprints{G.Tagliaferri (tagliaferri@merate.mi.astro.it)}
\institute{INAF-Osservatorio Astronomico di Brera, Via Bianchi 46, I-23807 Merate, Italy }
\date{Received ....; accepted ....}

\titlerunning{Observing Mkn 421 with XMM--Newton: the EPIC--PN point of view}
\authorrunning{Ravasio et al.\ }

\abstract{
We present three observations (four exposures) 
of Mkn 421 performed by XMM--Newton in the Autumn  of 2002,
concentrating on the EPIC--PN camera data. 
The  X--ray spectra were soft and steepening toward high energies. 
The source was  highly variable and the  hardness ratio plots 
displayed a clear harder--when--stronger correlation.
During  two complete flares the source showed strong spectral evolution:
a hardness ratio  and a time resolved spectral analysis
revealed both a clockwise and a counterclockwise  rotating  loop patterns,
suggesting the presence  of temporal lags between 
different energy bands variations.
We confirmed this result and estimated the delay amounts with 
a cross--correlation analysis performed on the single flares, 
discussing also variability patterns that could reproduce
the asymmetry seen in the cross-correlation function.
We verified our findings reproducing the two flares 
with analytical models. We obtained consistent results: during one flare, 
Mkn 421 displayed  soft lags, while in the other case it showed 
hard lags. In both cases, the entity of the delays increases
with the  energy difference between the compared light curves.
Finally,  we discussed the presence and the frequency dependence of  the temporal lags
as an effect of particle acceleration, cooling 
and escape timescales, showing that our data are consistent with this picture.
}
\maketitle
\section{Introduction}
The mostly accepted blazar models suggest that 
the multiwavelength continuum emission  is dominated by non--thermal
radiation from relativistic jets pointing close to the line of sight
(Urry \& Padovani 1995).
The Spectral Energy Distributions (SED) of blazars
are double--peaked, with a low energy  component peaking between
the IR and the X--ray band  and a high energy component 
peaking at GeV--TeV frequencies. 
While the first component is usually attributed to synchrotron emission,
the second one  is thought to be produced through inverse Compton scattering
between the electron population emitting via synchrotron mechanism and 
the synchrotron photon themselves (Maraschi, Ghisellini \& Celotti 1992)
or the photons of an external radiation field (Dermer \& Schlickeiser 1993; 
Sikora, Begelmann \& Rees 1994;
Ghisellini \& Madau 1996; Blazejowski et al. 2000).\\
Blazars are characterised by large and fast variability
on timescales even shorter than 1 hour (e.g. Mkn 421, Maraschi et al. 1999; 
BL Lac, Ravasio et al. 2002).
Since the highest energy part  of the  electron distribution
evolves more rapidly,  we expect  the variability events to be
energy--dependent, with the variations of the highest--energy  
section of the two SED components  leading those at smaller energies.
In the High Energy Peaked BL Lacs (HBLs), this behaviour
should be observable
mostly in the X--ray and in the TeV bands, where the 
synchrotron and the inverse Compton components peak, respectively.
In these bands, therefore, we should observe the largest 
and fastest flux variations, which could  be  characterised 
with observations even shorter than half a day.\\
Mkn 421 (z=0.031) is one of the brightest BL Lac objects in the UV 
and in the X--ray band and the first extragalactic source 
detected at TeV energies (Punch et al. 1992). 
It is classified as an HBL 
as its synchrotron peak lies close to the X--ray band.\\
It is very bright in the X--ray band, with the [2--10] keV
flux normally ranging in the 0.4--5$\times 10^{-10}$ erg cm$^{-2}$ s$^{-1}$
range, with the highest [2--10] keV flux 
($1.2\times 10^{-9}$ erg cm$^{-2}$ s$^{-1}$) 
recorded in May 2000 (Fossati 2001).
Because of its brightness, Mkn 421 has been the target
of almost every X--ray mission: the more recent campaigns 
were performed with ASCA (see e.g. Takahashi et al. 1996;
Takahashi et al. 2000), with {\it Beppo}SAX, which observed the source
intensively in May 1997, April 1998 and April 2000 
(Guainazzi et al. 1999; Fossati et al. 2000a; 
Fossati et al. 2000b; Malizia et al. 2000; Zhang 2002b)
and with XMM--Newton (Sembay et al. 2002; Brinkmann et al. 2003).\\
The  X--ray  behaviour of Mkn 421 is complex. 
Its historical  X--ray spectral shapes are usually soft 
above 1 keV and hardening toward lower energies.
Fossati et al. (2000b) fitted several {\it Beppo}SAX [0.1--10] keV spectra
taken in 1997 and 1998  with a  curved model.
They found that Mkn 421 spectra steepen continuously:
the  spectral indexes at 0.5 keV are hard ($\alpha \sim 0.7-1.2$)
and become softer toward higher energies
(at 10 keV, $\alpha\sim 1.5-2.3$). 
Fossati et al. (2000b)  also evidenced  that when the X--ray 
flux increases, the X--ray spectrum becomes harder 
and  the synchrotron peak  shifts to higher energies.
These results  were confirmed through a re--analysis of the 
historical {\it Beppo}SAX observations of Mkn 421 by Massaro et al. (2003b).
They  are also consistent with the results obtained from ASCA data
by Takahashi et al. (2000), 
which found spectral  indexes $\alpha\sim 1.4-1.8$ in the [2--7] keV band.
They  have been validated also by more recent observations
performed with XMM--Newton (Brinkmann et al. 2003).\\
Like other HBL blazars,
Mkn 421 is very variable both in the X--ray band and in the TeV band,
even on timescales of $\sim 20$ min (see e.g. Gaidos et al. 1996).
Several multiwavelength campaigns  were performed
to study the possible presence of lags between 
the TeV and the X--ray bands and the X--ray spectral evolution 
during  flares.
Thanks to simultaneous  {\it Beppo}SAX and Whipple observations
taken in 1998,  Maraschi et al. (1999) demonstrated that the X--ray and
TeV light curves are well correlated on timescales of hours
(and no lags are detectable).\\
Using ASCA, {\it Beppo}SAX and XMM--Newton data, several authors
reported the existence of temporal delays between the flux variations
at different X--ray energies in this 
and in other similar sources (see Sect. 4.2 for references). 
Their results were often controversial since the presence of 
soft lags, hard lags and no lags was claimed for different observation epochs.
XMM--Newton, thanks to its temporal resolution, higher throughput and particularly
to its gap--free observing modes can be particularly helpful in 
investigating the presence of temporal lags and their  
frequency dependence. \\
In this paper we will analyse 3 XMM--Newton observations (4 exposures)
taken in  Autumn 2002, concentrating on the EPIC--PN data:
in Section 2 we will present the observations and the reduction process.
Then we will describe  the spectral analysis performed
in the [0.6--10] keV range.
After having shown the light curves and the corresponding
hardness ratios, we will concentrate on two well defined
flares observed during two different nights.
On these two flares we performed a time--resolved spectral analysis
and a cross--correlation analysis, using also the 
discrete cross--correlation technique to check our results.
Finally a general discussion will be performed in Section~6 .
\begin{table*}
\begin{center}
\begin{tabular}{cccccc}
\hline
Revolution &  Obs. Id. & Start time & Obs. mode & Total exposure & Net exposure$^a$ \\
           &          &  (UT)      &           & ($\times 10^4$ s) & ($\times 10^4$ s) \\
\hline
0532 & 0136540301 & 2002--11--04 &  Prime Full & 2.39 & 1.29 \\
     &            & 01:07:55     &  Window     &      &  \\
0532 & 0136540401 & 2002--11--04 &  Prime Full & 2.39 & 1.28 \\
     &            & 08:04:39     &  Window     &      &            \\
0537 & 0136540701 & 2002--11--14 &  Prime Large & 7.15 & 3.77  \\
     &            & 00:07:35     &  Window      &      &            \\
0546 & 0136541001 & 2002--12--01 &  Timing      & 7.11 & 5.49 \\
     &            & 23:18:35     &              &      &             \\
\hline
\end{tabular}
\caption{Log of the observing campaign. $^a$: times referring to CCD--4.}
\label{tab1}
\end{center}
\end{table*}
\section{The XMM--Newton observation}
The XMM--Newton X--ray payload consists
of three Wolter type--1 telescopes, 
equipped with 3 CCD cameras (2 MOS and 1 PN) for X--ray imaging,
moderate resolution spectroscopy and X--ray photometry (EPIC).
Two of these telescopes (those carrying the MOS cameras)
are provided also with high resolution Reflection Grating Spectrometers
(RGS1 \& RGS2), deflecting half of the telescope beam.
In the further analysis
we will concentrate on the PN camera  which is less
affected by photon pile--up with respect to the MOS cameras
and which has better time resolution.
The PN camera consists of an array of 12 back--illuminated CCDs
 with a high sensitivity between 0.15 and 15 keV.\\
Mkn 421 was the target of an RGS and MOS calibration campaign during
the Autumn 2002, aimed at improving
the instrumental performances by lowering
the operating temperature. The source was observed during the nights 
of  November 4, November 14 and  December 1, 2002.
In Table \ref{tab1} we report
the log of the campaign referring to the PN camera:
the observations were performed in various operating modes,
characterised by different readout times.\\
We reduced the data using the XMM--Newton Science Analysis System
(SAS) 5.4.1 and the same calibration files used by the XMM--Newton
Survey Science Centre during  the standard Pipeline Processing.
For each observation, we extracted the light curves from
off--source circular regions, to check the presence
of high background periods, caused e.g. by solar flares. 
Because of the strong photon pile--up affecting  the inner source regions,
for the imaging observations
we  extracted the source events from annuli of radii 40'' and 1'20'', 
centered on the source  position. 
We chose these regions after having performed several tests
with the SAS task {\it epatplot} 
on different circular and annular regions and because, during the first two 
exposures (4 November), the inner source region was obscured by a square mask.
In the Timing mode observation, we extracted the source photons from a 
box 10 pixels RAW wide, centered on the source strip 
and extended all along the CCD.
In order to maximally avoid the pile--up effects, 
we accepted only single pixel events (PATTERN=0) with quality--flag=0.\\
The background event files were extracted from circular off--source regions
and from rectangular boxes away from the source strip for the 
imaging and Timing observations, respectively.
For the spectral analysis we used the canned response matrices
available at the XMM--Newton site and the ancillary files obtained with the SAS 5.4.1.
%
%
\begin{figure*}
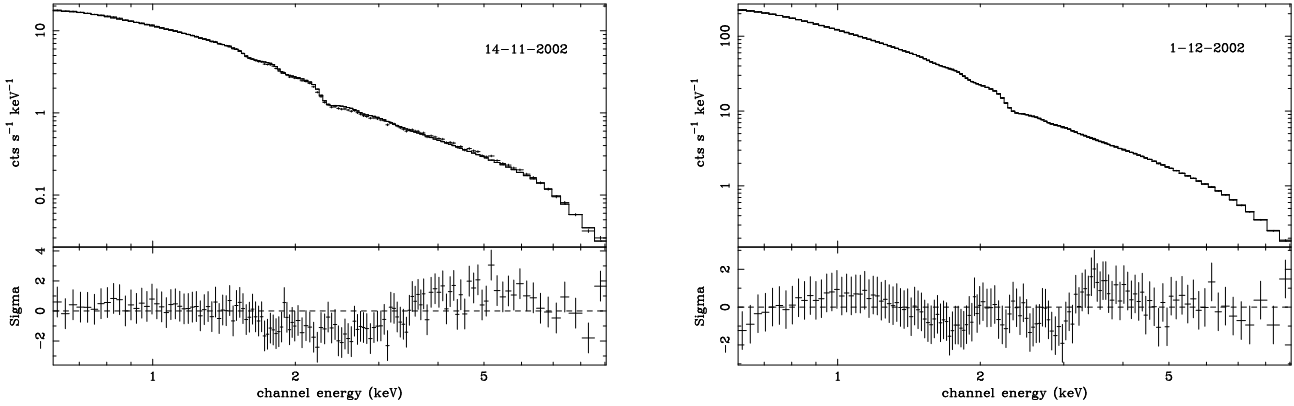

\hbox to \textwidth
{\centerline{
\vbox {\psfig{figure=figura1.ps,width=8cm,angle=-90}}
\hskip 1cm
\vbox{\psfig{figure=figura2.ps,width=8cm,angle=-90}}
\hfill
}}
\caption {Mkn 421 PN spectra of  November 14, 2002 (left) and 
of December 1$^{\rm st}$, 2002 (right). The first observation was performed 
in imaging mode, while the second was performed in Timing mode.
The spectra were both fitted with convex broken power--law models.
We added a 3\% systematic error to the imaging data and a 1.5\% systematic
error to the Timing mode data. The features in the residuals are caused
by the uncertainties in the calibration of the EPIC--PN response.}
\label{spec}
\end{figure*}
\section{Spectral analysis}
\begin{table*}
\begin{center}
\begin{tabular}{cccccccc}
\hline
Obs.Id. & $\alpha_1$ & E$_b$ & $\alpha_2$ & F$_{1 keV}$ & F$_{0.6-2 keV}^a$ & F$_{2-10 kev}^a$ & $\chi^2_r/dof$ \\
 & & (keV) & & ($\mu$Jy) & ($\times10^{-10}$) &  ($\times10^{-10}$) & \\
\hline
0136540301  & $1.53\pm0.01$ &  & & 123.8 & 3.34 & 2.25 & 1.21/125\\ 
0136540301  & $1.31^{+0.1}_{-0.16}$ & $1.00^{+0.25}_{-0.12}$ & $1.57^{+0.01}_{-0.02}$ & 128.3 & 3.35 & 2.21 & 0.98/123\\
\hline
0136540401 & $1.41\pm0.01$ & & & 154.8 & 4.20 & 3.31 & 1.40/125\\
0136540401 & $1.27^{+0.07}_{-0.09}$ & $1.14^{+0.24}_{-0.16}$ & $1.45\pm0.02$ & 157.1 & 4.21 & 3.25 & 1.20/123 \\
\hline
0136540701 & $1.15\pm0.01$ & & & 153.9 & 4.25 & 4.82 & 1.81/125\\
0136540701 & $1.13\pm0.01$ & $6.56^{+0.20}_{-0.16}$ & $2.26^{+0.17}_{-0.23}$ & 153.6 & 4.24 & 4.67 & 1.14/123\\
\hline
0136541001 & $1.535\pm0.003$ & & & 71.1 & 1.92 & 1.28 & 1.68/125\\
0136541001 & $ 1.521\pm0.004$ & $4.50^{+0.35}_{-0.27}$ & $1.70\pm0.03$ & 71.0 & 1.91 & 1.27 & 0.61/123\\ 
\hline
\multicolumn{8}{c}{Parabolic model}\\
\hline
Obs.Id. & $\alpha$ & & $\beta$ & F$_{1 keV}$ & F$_{0.6-2 keV}^a$ & F$_{2-10 kev}^a$ & $\chi^2_r/dof$ \\
 & &  & & ($\mu$Jy) & ($\times10^{-10}$) &  ($\times10^{-10}$) & \\
\hline
0136540301 & $1.45^{+0.03}_{-0.02}$ & & $0.16^{+0.03}_{-0.05}$ & 124.8 & 3.35 & 2.18 & 0.95/124\\
0136540401 & $1.34\pm0.02$ & & $0.14^{+0.03}_{-0.04}$ & 155.6 & 4.21 & 3.22 & 1.13/124\\
0136540701 & $1.16\pm0.02$ & & $0.02\pm0.01$ & 153.9 & 4.25 & 4.83 & 1.81/124 \\
0136541001 & $1.503^{+0.005}_{-0.003}$ & & $0.054\pm0.04$ & 71.2 & 1.92 & 1.27 & 1.24/124\\
\hline
\end{tabular}
\caption{Best--fit parameters of the absorbed power--law, broken 
power--law and  parabolic models.  We added a systematic error of 3\% 
to the  imaging mode data (exp. 0136540301, 0136540401 and 0136540701) 
and a systematic error of 1.5\% to the Timing mode data (0136541001).
$^a$: erg cm$^{-2}$ s$^{-1}$.}
\label{tab2}
\end{center}
\end{table*}
We concentrated the spectral analysis on the [0.6--10] keV 
energy range because of the large uncertainties in the PN detector
response below these frequencies (see e.g. Brinkmann et al. 2001; 
Brinkmann et al. 2003).
We rebinned the 4 PN spectra  in order to have 
a better Gaussian statistics and we fitted them with 
an absorbed power--law and  a broken power--law model. 
We always kept  the absorption parameter fixed  to the Galactic value
(N$_{\rm H} = 1.61\pm 0.1 \times 10^{20}$ cm$^{-2}$;
Lockman \& Savage 1995). 
To reduce the effects of the calibration uncertainties which are 
emphasised by  the very good statistics,  we added a systematic error 
of $3\%$ to the data taken in imaging mode
and of $1.5\%$ to the data taken in Timing mode.
We decided to use a lower systematic error for the December 1$^{\rm st}$ data
since adding a 3\% error greatly overestimates the uncertainties:
the $\chi^2_r$ of each fit would result smaller than 0.5 (the Timing data have
higher intrinsic background and therefore a smaller systematic error is needed).
The best--fit spectral parameters for each observation
are reported in Table \ref{tab2}.\\
All the PN spectra are better fitted by a convex broken power--law model 
than by a simple power--law model: in each case, the F--test  probability 
of improving the quality of the fit is $> 99.9\%$. The X--ray spectra
of Mkn 421 become systematically  softer toward higher  energies, 
confirming the results of several previous observations
(see e.g Fossati et al. 2000b; Brinkmann et al. 2003).\\
In Fig.\ref{spec} we plot two PN spectra fitted by a broken 
power--law model: the November 14 spectrum was taken in imaging mode
while the December 1$^{\rm st}$ spectrum was collected in Timing mode.\\
We tried  to reproduce the spectra also with a curved 
model which can account for the progressive steepening.
We used the logarithmic parabolic model described 
by Massaro et al. (2003a, b), which should provide a reasonable representation
of the wide band spectral distribution for the low energy component of 
blazars:
\begin{equation}
F(E) = K (E/E_1)^{-(a+bLog(E/E_1))}
\end{equation}
where, in our computations $E_1 = 1$ keV.\\
This model fits the  November 4 data well: it provides 
a similar or lower $\chi^2_r$ than the broken power--law model.
On the contrary, the November 14 and the December 1$^{\rm st}$ data
are better represented by  broken power--law models.  
In Table \ref{tab2} we report the best--fit parameters of all the described 
models. 
\section{Light curves}
\begin{figure}
\centerline{
\vbox{
\psfig{figure=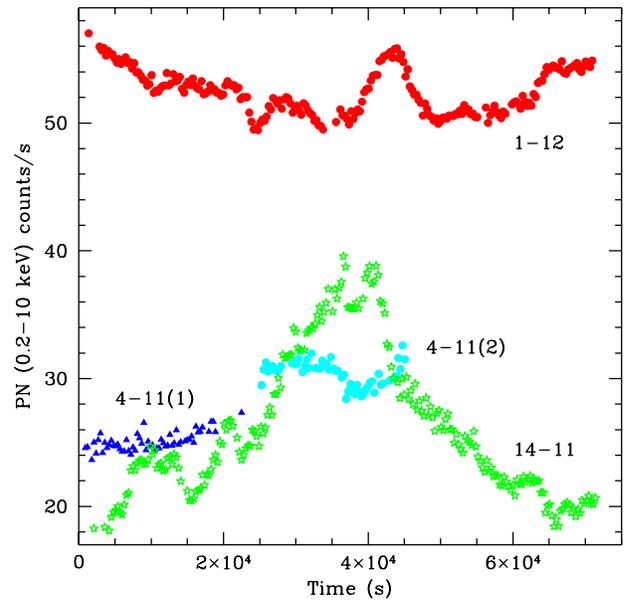,width=8.5cm}
}}
\caption{EPIC--PN [0.2--10] keV light curves of the four 
observational periods. The second exposure 
of November 4 is plotted as the continuation of the first.
For plotting reasons, the  December 1$^{\rm st}$ light curve,
obtained in Timing mode, is rescaled down by a factor of 10 (see text).
For clarity we do not plot the error bars (that in any case 
are comparable with the symbol sizes).}
\label{lcurves} 
\end{figure}

In Fig. \ref{lcurves}, we plot the 300 sec binned  light curves
of Mkn 421 in the [0.2--10] keV range.
In the rest of the paper we will exclude from the analysis all the
temporal bins with less than 30\% of effective exposure
(i.e. points for which the data are collected for less than 30\% of the time).
For plotting  purposes, in Fig. \ref{lcurves}
the count rate of December 1$^{\rm st}$ 
was divided by a factor of 10. Note, however, that the larger count
rate during this observation is not related to a higher source flux
(see Table \ref{tab2})
but to a different operating mode. Since this observation 
was performed in Timing mode we were not forced to discard  photons
to avoid pile--up effects. From the figure, we can note that:
\begin{itemize}
\item November 4: during the first exposure, 
the source flux  increases slowly.
Between the first and the second run the flux sudden rises by $\sim 25\%$
then we observe it decreasing by $\sim 15\%$; finally the source rebrightens
to the  previous maximum level.
\item November 14: a large and complete flare 
lasting a few hours is present in the light curve:
the [0.2--10] keV count rate doubles and fades to previous values.
\item December 1: very small features  are  present in the light curve,
but a  well defined small flare can be observed
after about half observation and lasting $\sim 4$ hrs, 
with a flux increase of $\sim 10\%$. 
Since during this night the PN was operating in  Timing mode,
we were able  to perform a detailed temporal analysis also on
this small feature.
\end{itemize}
\begin{figure}[!ht!]
\centerline{
\vbox{
\psfig{figure=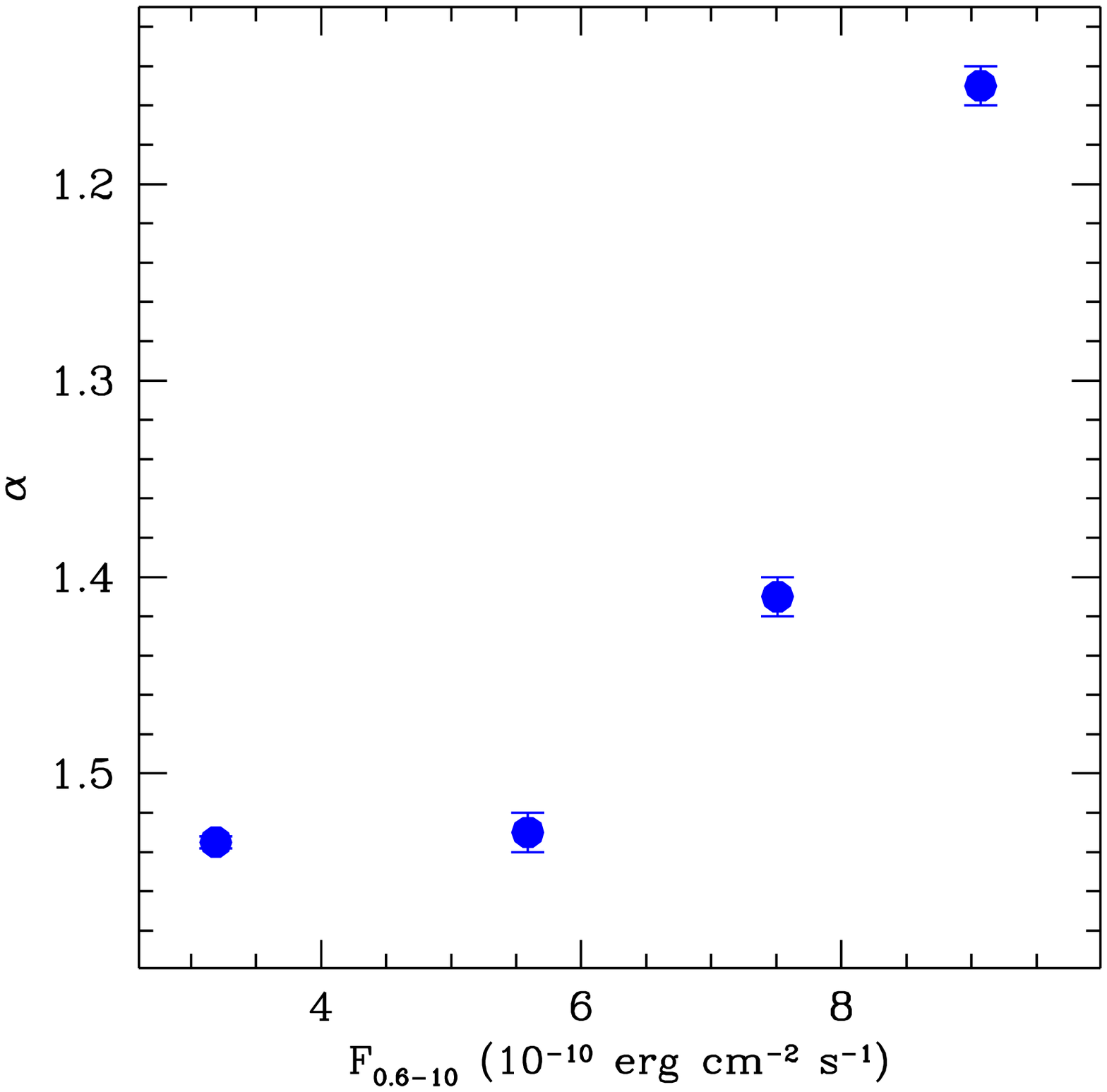,width=8.5cm}
}}
\caption{The best--fit  power--law model spectral indexes
versus the [0.6--10] keV flux. The source is harder when 
the [0.6--10] keV flux is higher.}
\label{alpha-f} 
\end{figure}  
In order to quantitatively estimate the  source variability,
we  calculated the normalized excess variance of the 300 s
binned light curves  in different  energy bands:
[0.2--0.8] keV, [0.8--2.4] keV and [2.4--10] keV.
We report our results in Table \ref{tab3}.
According to Table \ref{tab2} and Table \ref{tab3},
there is a trend indicating that the source is more variable
while in a higher state of activity. We also found that
the source is systematically more variable toward higher energies.
This is not surprising in the framework of a standard leptonic model,
(see e.g. Ghisellini et al. 1999), since the X--ray spectrum
of an HBL should be produced via synchrotron emission. 
Harder X--rays are therefore produced by more energetic particles
with smaller cooling timescales.
Furthermore, since our hard X--ray  spectra 
are systematically steeper than the soft X--ray spectra, a small
change in the shape of the injected particle distribution
will produce greater variations toward higher energies.
\subsection{Hardness ratios}
\begin{figure*}
\hbox to \textwidth
{\centerline{
\vbox {\psfig{figure=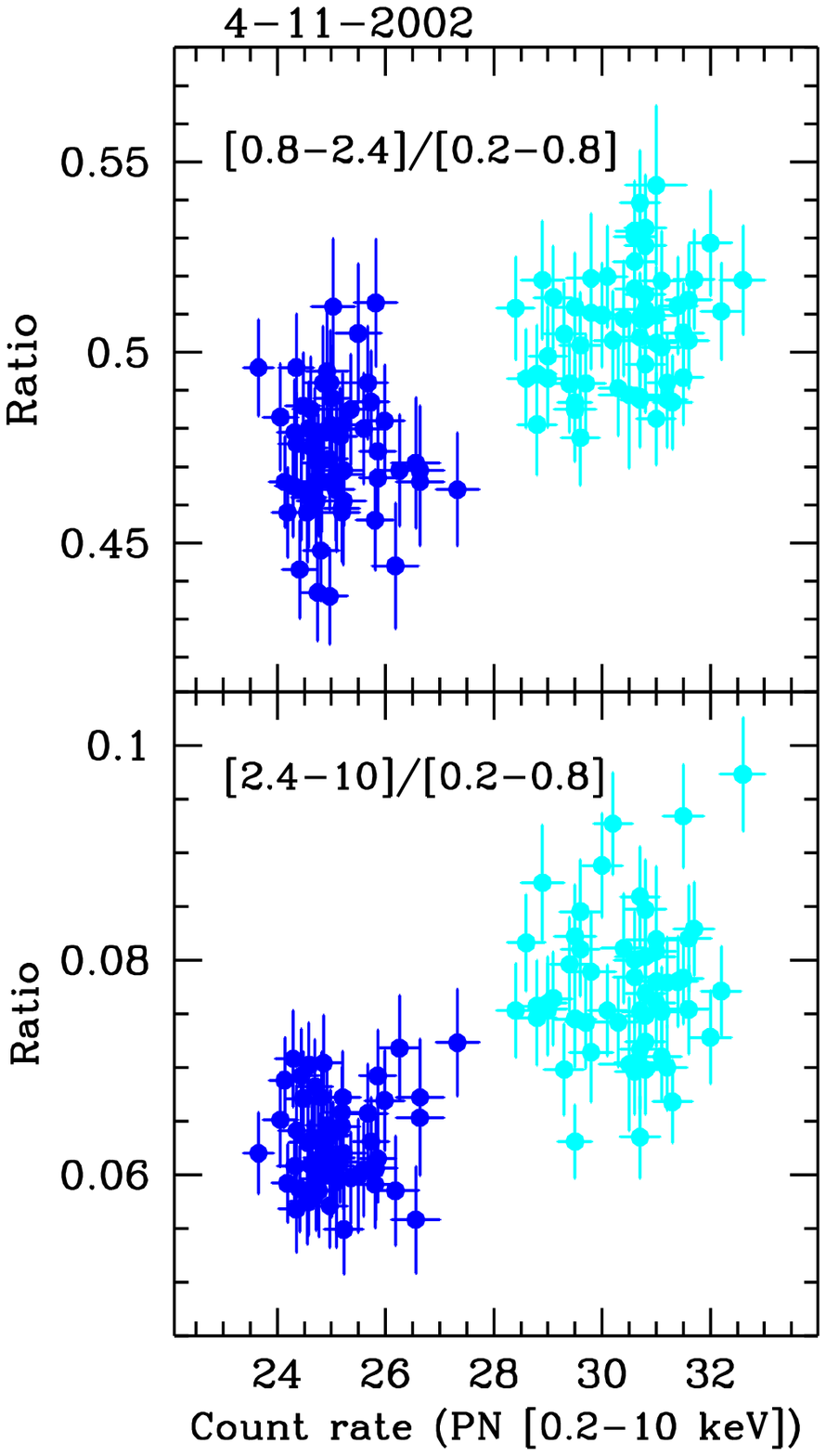,width=6cm}}
\hskip 0.cm
\vbox{\psfig{figure=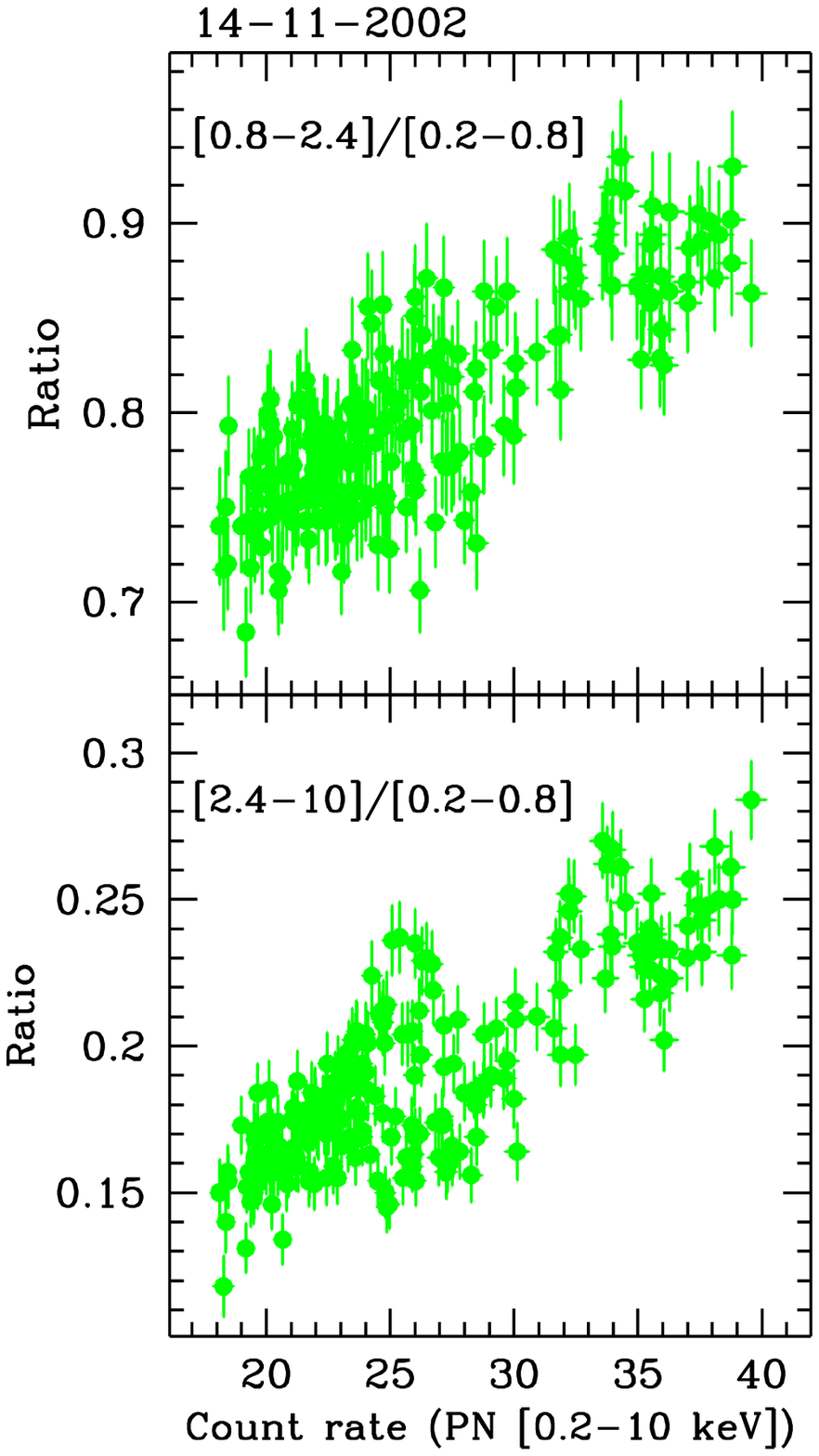,width=6cm}}
\hskip 0.cm
\vbox{\psfig{figure=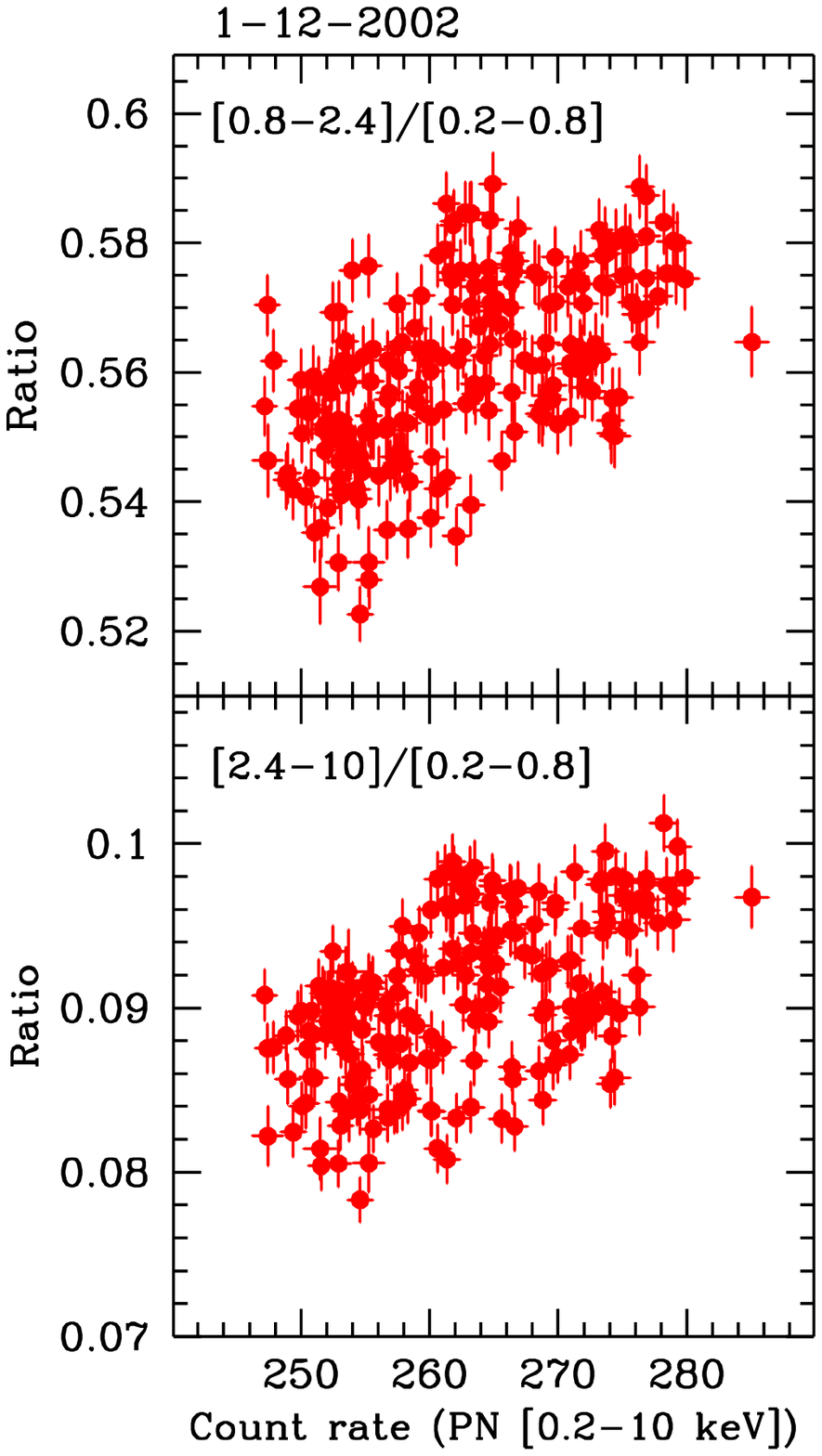,width=6cm}}
\hfill
}}
\caption {Upper panels: PN[0.8--2.4] keV / PN[0.2--0.8] keV ratios
versus the [0.2--10] keV count rates. 
Lower panels: PN[2.4--10] keV / PN[0.2--0.8] keV ratios
versus the [0.2--10] keV count rates. In the November 4 plot, 
we represent in dark grey the data of the first exposure 
and in light grey the data of the second exposure.}
\label{flux-ratio}
\end{figure*}

\begin{figure*}
\begin{center}
\hbox to \textwidth
{\centerline{
\vbox {\psfig{figure=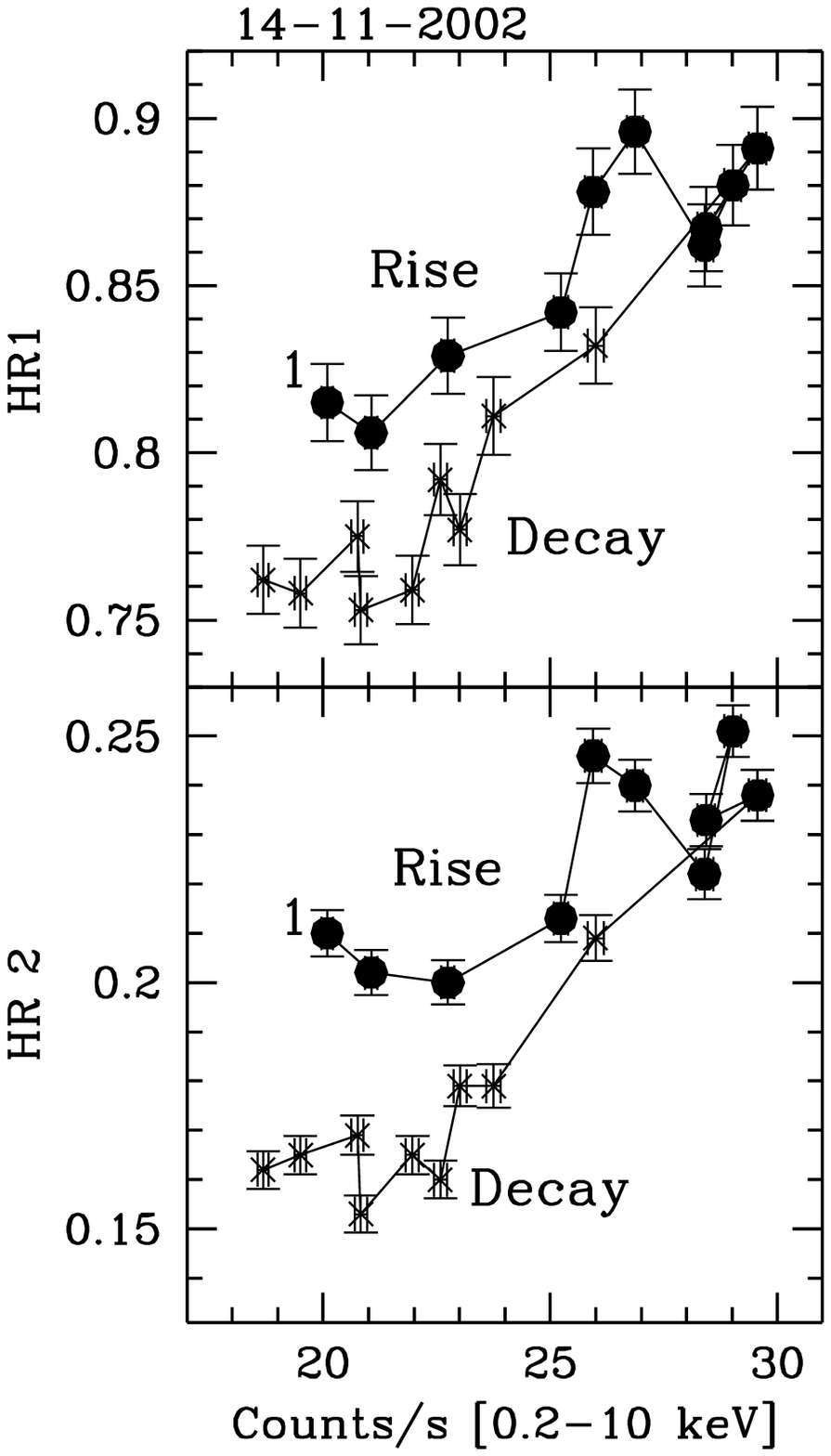,width=7cm}}
\hskip 1.5cm
\vbox{\psfig{figure=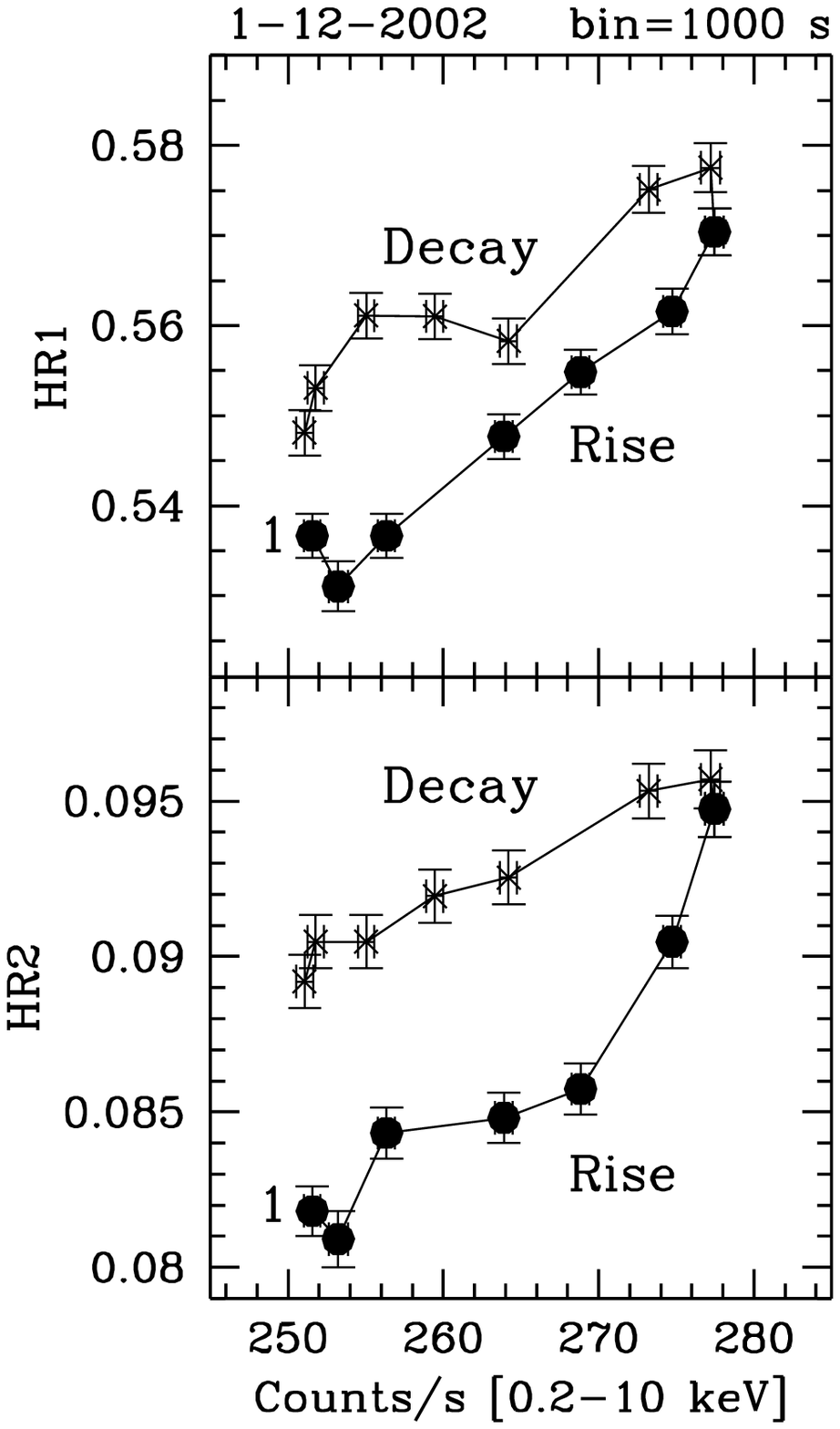,width=7cm}}
\hfill
}}
\caption {Hardness ratio/[0.2--10] keV count rate correlations for the 
flaring sections  of the November 14 and  December 1$^{\rm st}$
light curves (see text).
HR1=[0.8--2.4 keV]/[0.2--0.8] keV; HR2=[2.4--10] keV/[0.2--0.8] keV.
The rising phase data are plotted with circles, while the decaying phase data are
plotted with crosses. Each  temporal sequence starts from the data point marked with ``1''.
}
\label{clock}
\end{center}
\end{figure*}

\begin{table*}
\begin{center}
\begin{tabular}{c|cccc}
\hline
Obs. period & \multicolumn{4}{c}{$\sigma^2_{rms}$} \\
            & \multicolumn{4}{c}{($\times 10^{-3}$)}  \\
            & [0.2--0.8] keV & [0.8--2.4] keV & [2.4--10] keV & [0.2--10] keV \\
\hline
4--11(1) & $0.33\pm0.15$ & $0.39\pm0.30$ & $2.53\pm2.00$ & $0.35\pm0.15$ \\
4--11(2) & $0.76\pm 0.19$ & $1.23\pm0.40$ & $3.59\pm1.91$ & $0.78\pm0.17$ \\
14--11    & $21.14\pm 2.24$ & $40.96\pm4.33$ & $86.95\pm 10.04$ & $32.76\pm3.46$ \\
1--12 & $0.77\pm0.08$ & $1.73\pm0.16$ & $4.02\pm0.43$ & $1.07\pm0.10$ \\
\hline
\end{tabular}
\caption{The excess variances of the 300 s binned light curves.}
\label{tab3}
\end{center}
\end{table*}
With these data obtained weeks apart,
we have the possibility to check the spectral behaviour of the source 
both on long and on short timescales.\\
\begin{figure*}
\hbox to \textwidth
{\centerline{
\vbox {\psfig{figure=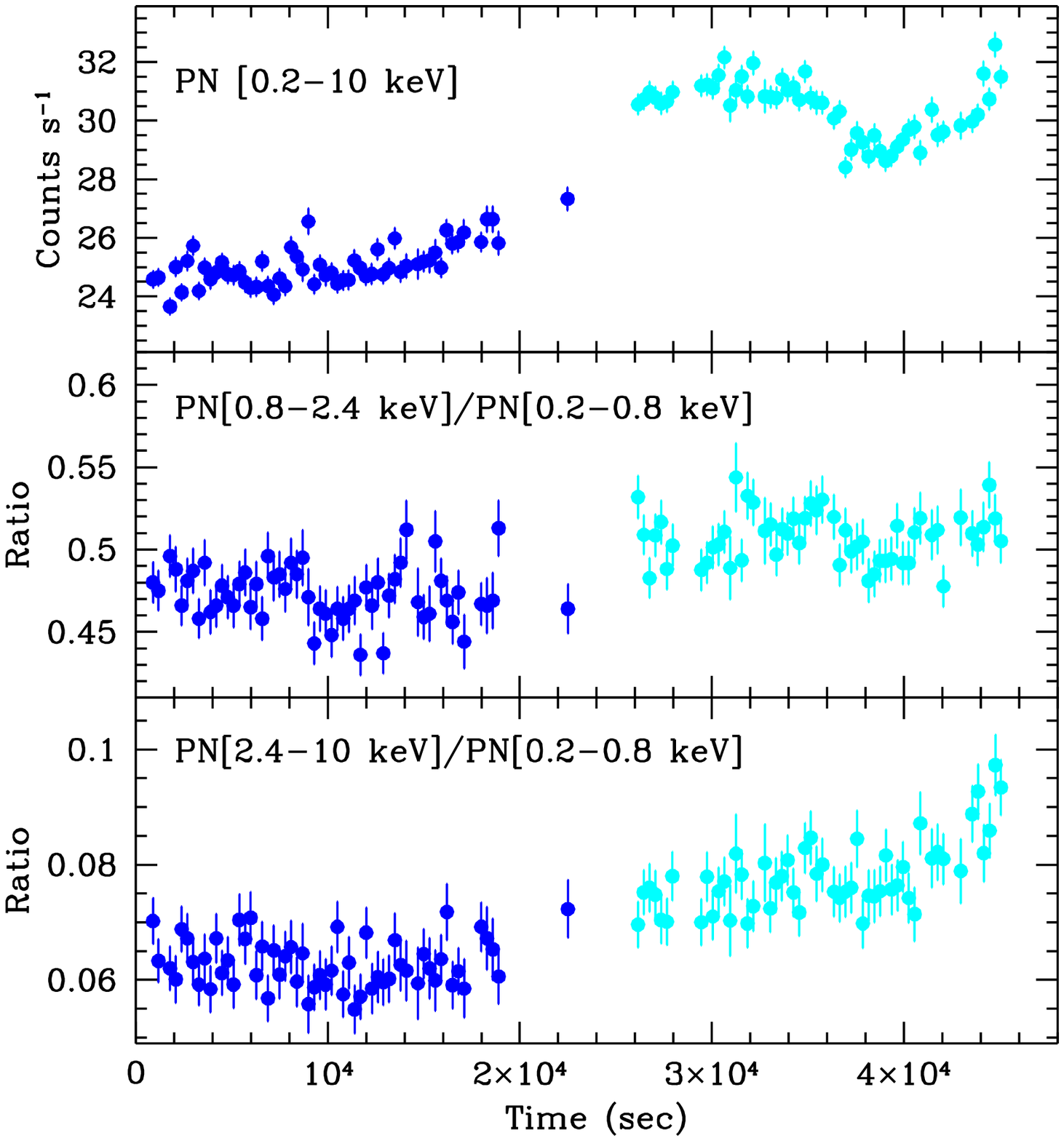,width=8.5cm}}
\hskip 1.cm
\vbox{\psfig{figure=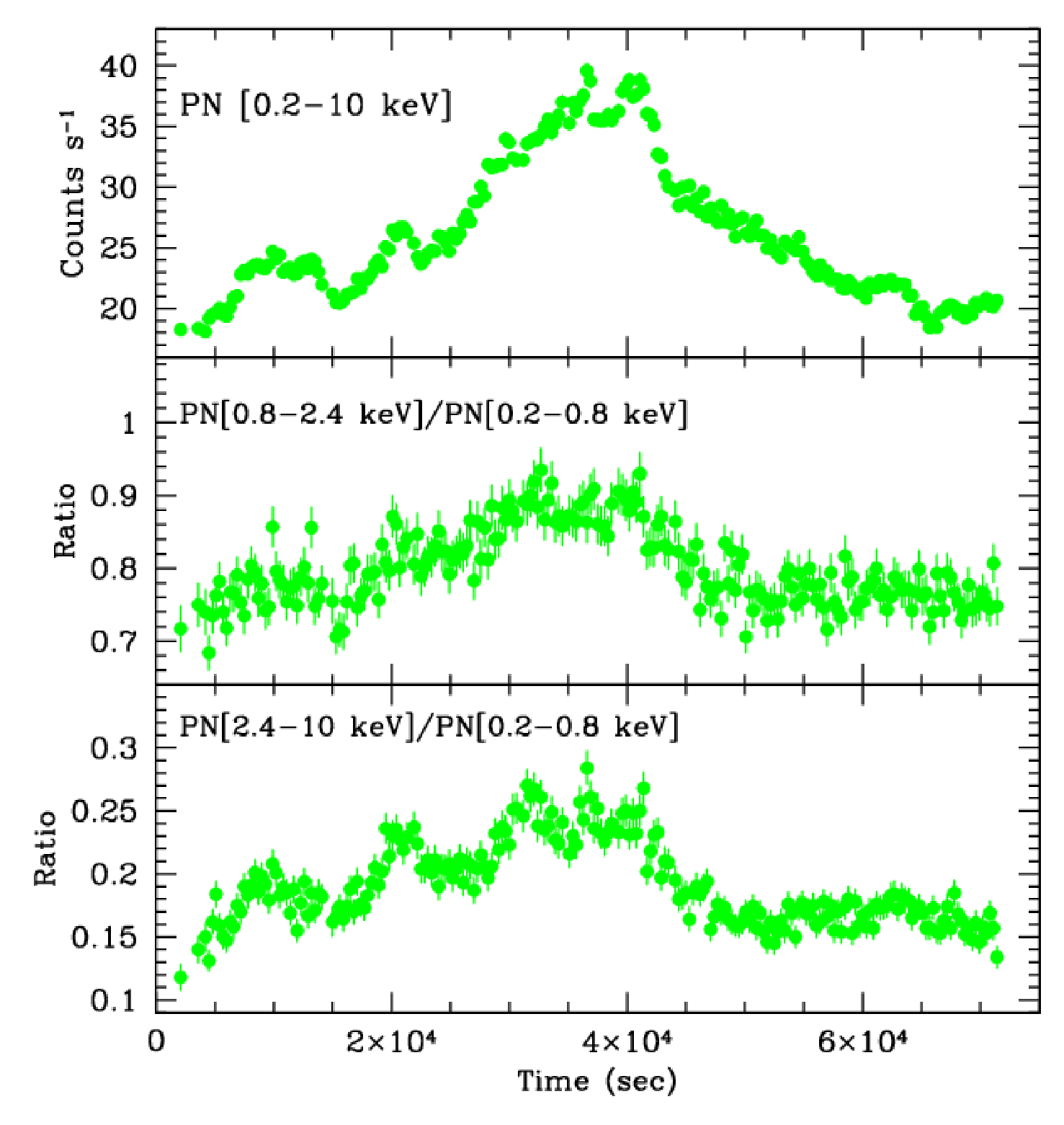,width=8.5cm}}
}}
\caption{ Left picture: November 4  EPIC--PN observation.
Right picture November 14 EPIC--PN observation. 
The upper panel reports the PN [0.2--10] keV light curve, 
the mid panel the PN[0.8--2.4] keV/PN[0.2--0.8] keV hardness ratio
and the lower panel the PN[2.4--10] keV/PN[0.2--0.8] keV hardness ratio.
We plot together the two observations
of  November 4 (0136540301 and 0136540401).}
\label{hard-curve-1} 
\end{figure*}  
\begin{figure}[!h!]
\hbox to \textwidth
{\centerline{
\vbox {\psfig{figure=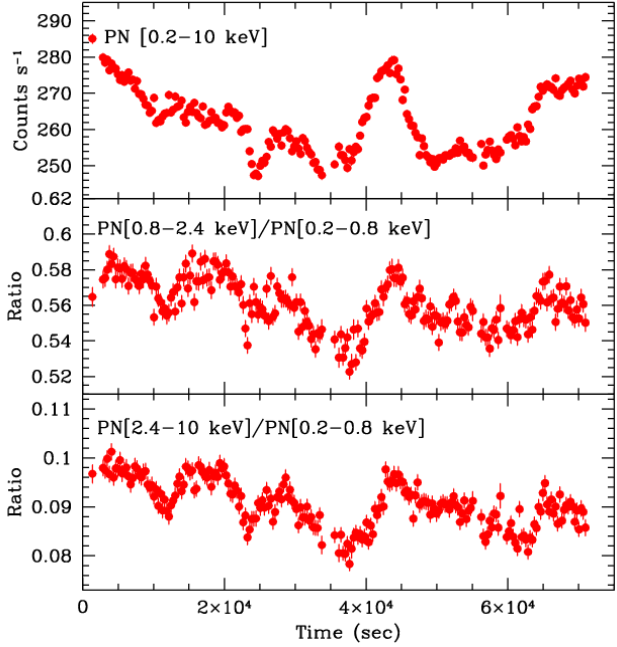,width=8.5cm}}
}}
\caption{December 1$^{\rm st}$ EPIC--PN observation.
The upper panels report the PN [0.2--10] keV 300  light curves, 
the mid panels the PN[0.8--2.4] keV/PN[0.2--0.8] keV hardness ratios
and the low panels the PN[2.4--10] keV/PN[0.2--0.8] keV hardness ratios.}
\label{hard-curve-3} 
\end{figure}       
To study the long term trend, we compared 
the best--fit spectral indexes of the absorbed power--law model
(which still provide  reasonable fits to the data)
to the total [0.6--10] keV fluxes reported in Table \ref{tab2}. 
We found that the X--ray  spectra of Mkn 421 are harder when 
the fluxes are stronger (see Fig.\ref{alpha-f}),
as was already observed during other X--ray campaigns
on Mkn 421  (see e.g. Fossati et al. 2000b; Sembay et al. 2002;
Brinkmann et al. 2003)
as well as on other similar sources (e.g. Mkn 501, Pian et al. 1998;
1ES 2344+514, Giommi et al. 2000; PKS 2155-304, Zhang et al. 2002a).\\
We checked this behaviour also on smaller timescales analysing
the hardness ratios of  light curves at different energies.\\
In Fig. \ref{hard-curve-1} and Fig. \ref{hard-curve-3} 
we plot the total [0.2--10] keV light curves (top panels), 
together with the [0.8--2.4] keV/[0.2--0.8] keV (mid panels) 
and the [2.4--10] keV/[0.2--0.8] keV hardness ratio
(bottom panels) for each observing night.\\
In  Fig. \ref{hard-curve-1} and Fig. \ref{hard-curve-3}
it is clear that the hardness ratios are correlated
with the [0.2--10] keV count rates: when the total flux increases
the spectra become harder and conversely. 
This is verified both for  long term variations 
(e.g. the slow flux increase observed during the whole
November 4 observation,  $\sim 4\times10^4$ s), 
for short term variations (e.g. the small flare observed
on December  1$^{\rm st}$,  $\sim  10^4$ s), for large events 
(e.g. the flare of November 14, flux variation $\gtrsim 100\%$) 
and for smaller events e.g. the same December flare ($\sim 10\%$).\\
This harder--when--stronger behaviour is also shown in
the hardness ratio vs [0.2--10] keV count rate plots 
(see Fig. \ref{flux-ratio}).
The hardness ratios are correlated with the [0.2--10] keV flux: Mkn 421
becomes harder as the [0.2--10] keV flux increases.
The null--correlation probability is always $< 10^{-10}$.\\
To investigate  the harder--when--stronger behaviour in more detail
we concentrated on the November 14
and December 1$^{\rm st}$ observations, where two complete flares,
different in amplitude and timescales,  were detected.\\
Studying these two events,
we investigated the spectral shape evolution during a whole 
flare, obtaining information on the particle acceleration/injection timescales
(rising phase of the flare), on the cooling timescale (decaying section
of the flare) and on  the region geometry.\\
During the  November 14 observation, the [0.2--10] keV counts 
increased by a factor larger than 2 
and then decreased to the initial level in a total 
time of $\sim 7\times 10^4$ s.
To avoid confusion  caused by the small flares 
at the beginning and at the end of the observation, 
we excluded  from the  analysis the first $2.5\times 10^4$ s 
and the last 5000 s.
For the observation of December 1$^{\rm st}$, we analysed the small flare 
(lasting $1.4\times10^4$ s) 
detected  $\sim 4\times10^4$ s after the beginning of the observation.\\
We  rebinned these  sections of the
light curves in 2000 s and 1000 s bins, respectively.
In Fig. \ref{clock}, we plot the hardness ratios HR1
([0.8--2.4] keV/[0.2--0.8] keV) and HR2
([2.4--10] keV/[0.2--0.8] keV) as a function of the 
total [0.2--10] keV count rates.
The rising phase data are plotted as filled circles and the
decaying phase data as crosses.
Besides  the above mentioned harder--when--stronger  trend, 
in Fig. \ref{clock} we note a substantial different behaviour
during the two flares:
in the November 14 rising phase (circle symbols), the source 
is slightly harder than in the decaying phase  
(cross symbols), forming  clockwise loop patterns.
In the December 1$^{\rm st}$ flare, the source behaves
in the opposite way:
during the rising phase  the source is systematically  softer
than in  the decaying phase,
forming a counterclockwise loop pattern.\\
In order to check the reality of these particular patterns,
we performed a time resolved spectral analysis of the two  flares.
\subsection{Time resolved spectral analysis} 
\begin{figure*}
\begin{center}
\hbox to \textwidth
{\centerline{
\vbox {\psfig{figure=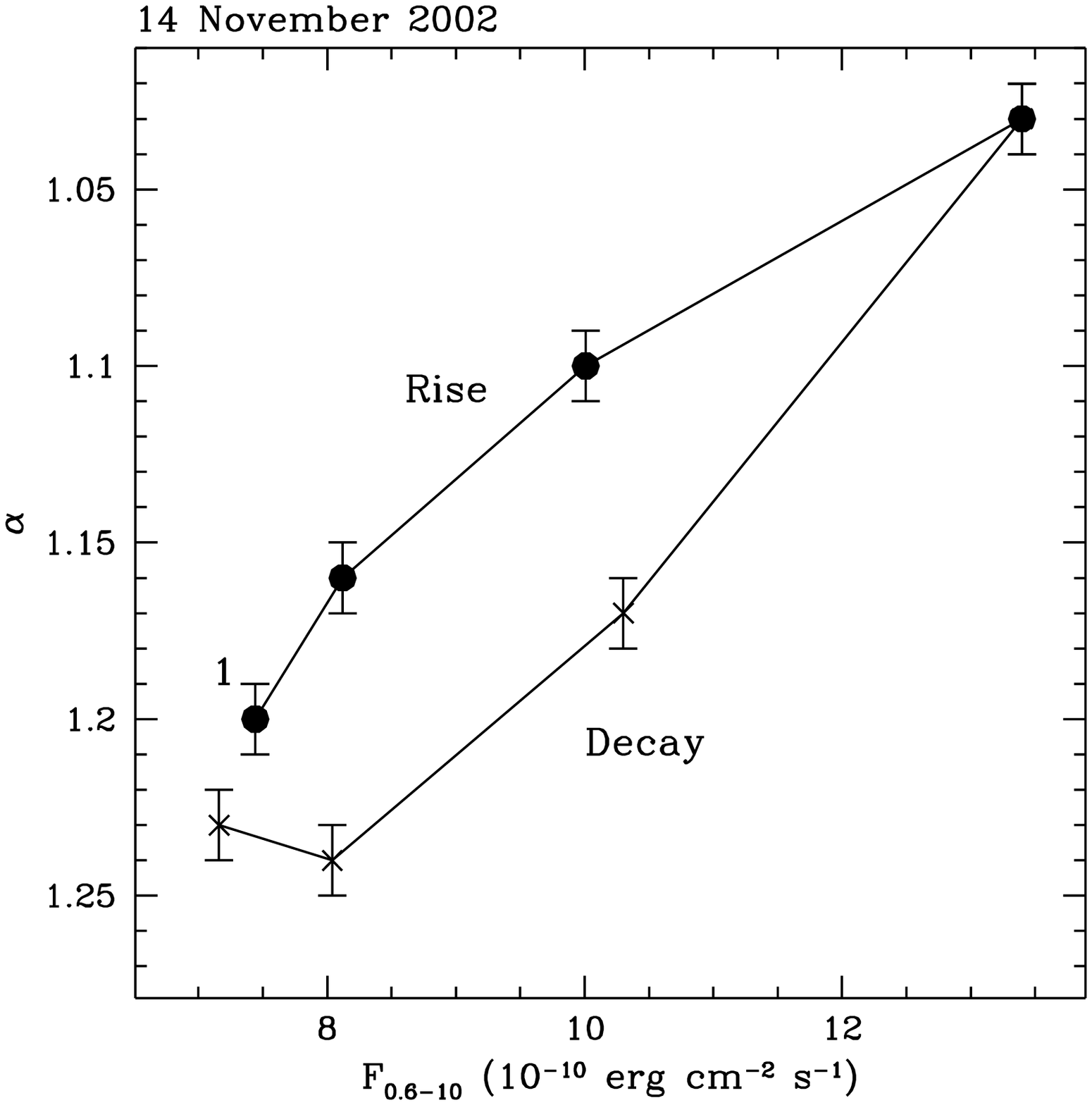,width=7.5cm}}
\hskip 1.5cm
\vbox{\psfig{figure=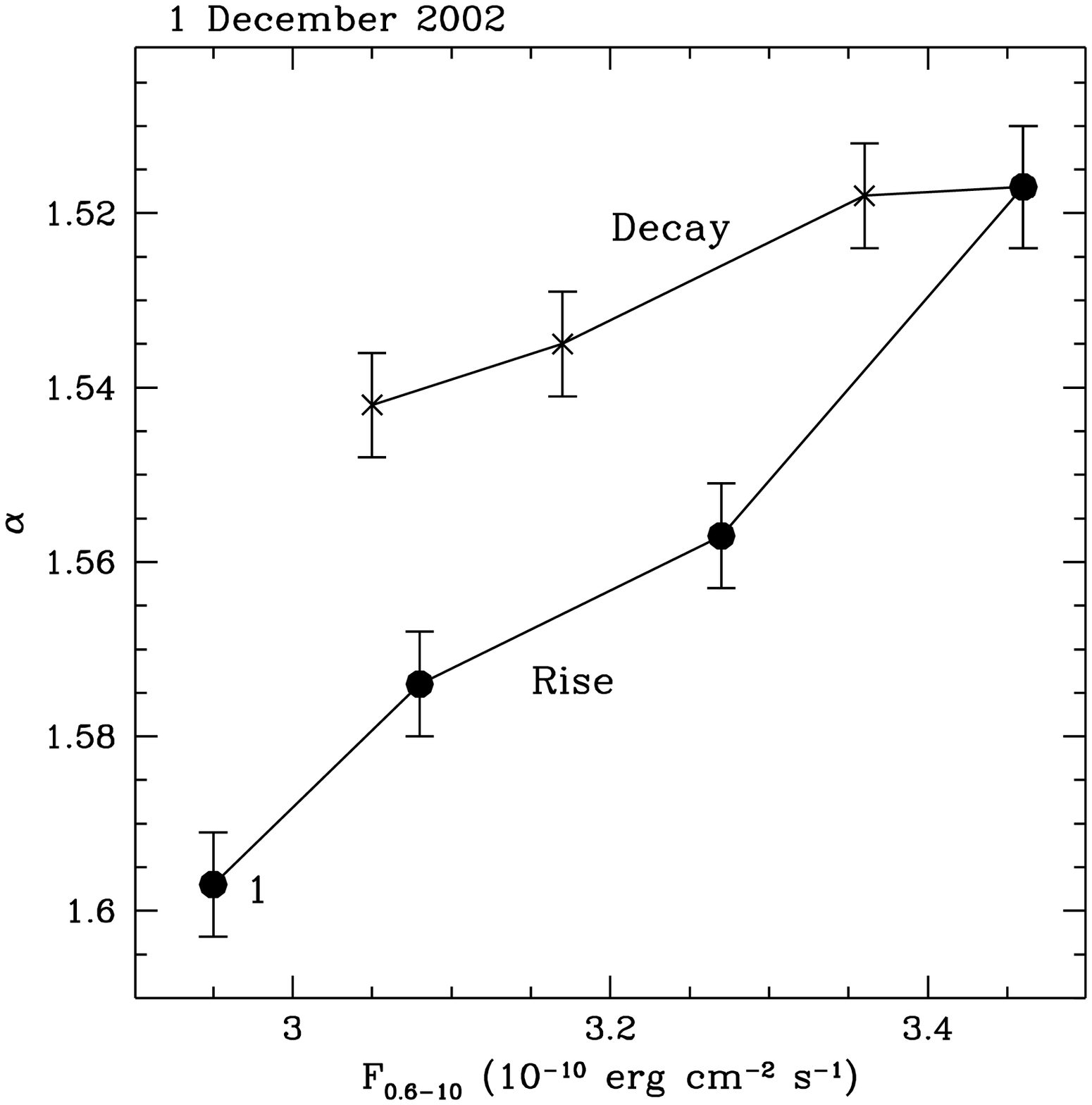,width=7.5cm}}
\hfill
}}
\caption {We plot the best--fit spectral indexes $\alpha$
versus the [0.6--10] keV fluxes of the spectra extracted from
seven sections of the November 14 exposure
and from seven sections of the small flare occurred during 
the December 1$^{\rm st}$ observation.
The spectra are modelled with an absorbed power law model.
We plot  the rising phase data with filled circles 
and the decaying phase data with crosses.
The points marked with ``1'' are the first
spectra of the temporal sequence.}
\label{flux-alpha}
\end{center}
\end{figure*}

We divided the  November 14 observation
in seven 10 ks sections and extracted the corresponding spectra.
The extraction of the data and the filtering processes were performed 
as described in Section 2.
We fitted each [0.6--10] keV spectrum with an absorbed  power--law model
keeping the absorption parameter fixed to the Galactic value.
Because of the lower statistic, this model provides
already a good representation of these spectra.\\
We performed the same analysis on the small flare
of December 1$^{\rm st}$. Since this observation 
was carried out in Timing mode, we have enough photon counts to 
split the short flare ($\sim 1.4 \times 10^4$ s) in seven 2000 s
sections and to extract well defined spectra from each of them. 
In Table \ref{tab4} we report the best--fit spectral parameters
for each temporal section of both flares.\\
\begin{table*}
\begin{center}
\begin{tabular}{ccccc}
\hline
Section & $\alpha$ & k$^a$ & F$_{0.6-10 keV}$$^b$ & $\chi^2_r/d.o.f.$ \\
        &          &  & ($\times 10^{-10}$) & \\
\hline
\multicolumn{5}{c}{14 November 2002, Obs. Id. 0136540701}\\
\hline
1  & $1.20 \pm 0.01$ & $0.199\pm 0.002$ & 7.44 & 1.28/125\\
2  & $1.16^{+0.01}_{-0.02}$ & $0.211\pm 0.001$ & 8.12 & 1.73/125\\
3  & $1.10\pm0.01$ & $0.250\pm 0.002$ & 10.1 & 1.70/125\\
4  & $1.03\pm0.01$ & $0.312\pm0.002$ & 13.4 & 1.72/125\\
5  & $1.17\pm0.01$ & $0.269^{+0.001}_{-0.002}$ & 10.3 & 1.35/125 \\
6  & $1.24\pm0.01$ & $0.222^{+0.001}_{-0.002}$ & 8.04 & 1.39/125\\
7  & $1.23\pm0.01$ & $0.197\pm0.002$ & 7.16 & 1.31/125\\
\hline
\multicolumn{5}{c}{1 December 2002, Obs. Id. 0136541001}\\
\hline
1 & $1.597\pm 0.006$ & $0.1027\pm0.0004$ & 2.95 & 1.02/977\\
2 & $1.574^{+0.006}_{-0.007}$ & $0.1057^{+0.0003}_{-0.0004}$ & 3.08 & 0.96/998\\
3 & $1.557\pm 0.006$ & $0.1113^{+0.0003}_{-0.0004}$ & 3.27 & 1.15/1029\\
4 & $1.517^{+0.007}_{-0.006}$ & $0.115^{+0.0003}_{-0.0005}$ & 3.46 & 1.02/1026\\
5 & $1.518\pm0.006$ & $0.1118^{+0.0003}_{-0.0004}$ & 3.36 & 1.09/1049\\
6 & $1.535\pm0.006$ & $0.1066\pm0.0004$ & 3.17 & 1.06/1038\\
7 & $1.542^{+0.006}_{-0.007}$ & $0.1029\pm0.0004$ & 3.05 & 1.07/1033\\
\hline
\end{tabular}
\caption{ Best--fit parameters of the seven spectra extracted from the 
November 14, 2002 observation and from the flaring section of the
December 1$^{\rm st}$, 2002  observation, modelled
with an absorbed power law.
$^a$: power law normalization (cts cm$^{-2}$ s$^{-1}$ keV$^{-1}$);
$^b$: erg cm$^{-2}$ s$^{-1}$.}
\label{tab4}
\end{center}
\end{table*}
The spectra become harder as the [0.6--10] keV flux increases
and then soften to the initial shape as the source fades.
This is shown also in Fig. \ref{flux-alpha}  where
we plot the best--fit spectral indexes versus the [0.6--10] keV flux.
Fig. \ref{flux-alpha} also  shows the same clockwise (November 14)
and  counterclockwise loop patterns (December 1$^{\rm st}$) 
obtained from the hardness ratio analysis.\\
These characteristic trends were already observed during
previous campaigns on Mkn 421:
performing a temporally resolved spectral analysis on ASCA data,
Takahashi et al. (1996)  were the first to observe a clockwise loop pattern 
which was interpreted as the signature of a soft lag ($\sim 1$h), 
i.e. hard X--ray variations leading soft X--ray variations.
Fossati et al. (2000b), instead, were the first to 
find a counterclockwise loop pattern in a Mkn 421 
flare observed by {\it Beppo}SAX, 
which they explained as the sign of a hard lag ($ \sim 2-3$ h), i.e.
soft X--ray variations leading hard X--ray variations.
They confirmed this evidence performing also a discrete cross--correlation
analysis. Using the same technique on different sections
of an ASCA light curve of April 1998, Takahashi et al. (2000)
found evidences of soft ($\sim 2000$ s), hard ($\sim 3400$ s) and of no lags.\\
Performing a discrete cross--correlation analysis on  4
XMM--Newton orbits, Sembay et al. (2002)
did not found lags. They  suggested that the previous detections
were caused by systematic errors induced by gaps in the on--source time
of low Earth orbit satellites such as {\it Beppo}SAX and ASCA.
Brinkmann et al. (2003) re--analysed the same and other XMM--Newton data,
dividing the light curves in sub--sections characterised 
by single flaring events.
In different sections of the light curves, they found  
soft and  hard lags as well as  no lags, confirming
the extremely complex behaviour of the source.\\
Similar  behaviours were detected also in other sources, 
such as PKS 2155-304  (Kataoka et al. 2000; Zhang et al. 2002a)
or BL Lacertae (B\"ottcher et al. 2003).\\
\section{Delay determination}
To check the presence and to estimate the amount of the temporal delays 
between flux variations in different energy bands, we performed
two different kind of analysis.
We concentrated  on the two main variability features observed
in the 4 EPIC--PN exposures, i.e. the large and structured 
flare seen on  November 14, covering the whole XMM--Newton 
observation, and the small flare observed during
the December 1$^{\rm st}$ observation.\\
The  delay between two light curves is usually estimated by fitting 
the central peak of their cross--correlation function (CCF)
with a Gaussian profile and taking the centroid position as the delay value. 
This technique, however, must be used cautiously: 
while it works properly for single, smooth and symmetrical
flares, it can give unreal results when  
used on structured or asymmetrical light curves.
In these cases, the CCF shape is deformed  
and the best--fit position of the Gaussian centroid will
roughly be a weighted  average of the delays between  the
several components or an index of the light curves asymmetry.
Since our flares display complex shapes, in order to 
avoid confusion and wrong delay estimations,
we fitted the CCF peaks with an asymmetrical model
(e.g. Brinkmann et al. 2003), 
and checked the results by fitting the light curves with analytical models,
to disentangle the various subcomponents.
Comparing the locations of the maxima and  the minima
we obtained independent  delay estimations.
In the following Sections we will describe in detail these 
techniques and the results obtained.
\subsection{Cross--correlation analysis}
\begin{figure}
\centerline{
\vbox{
\psfig{figure=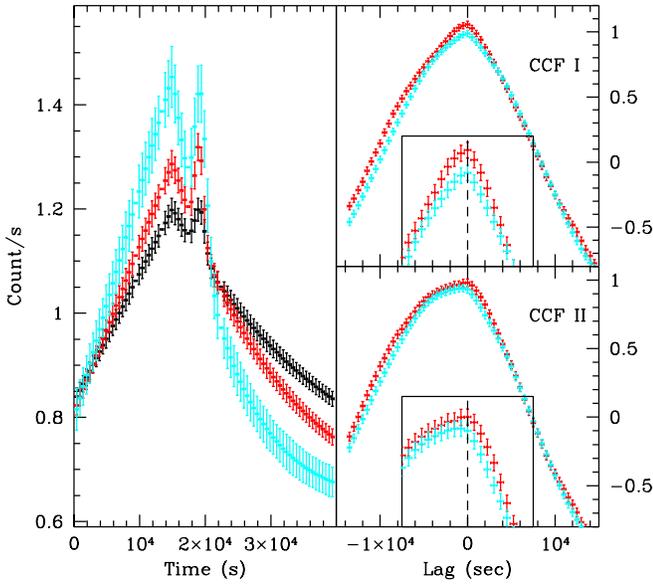,width=9cm}
}}
\caption{ 
{\it Left panel}: 500 s binned light curves generated from the
best--fit models to the main flare of November 14 (see text). 
We show the [0.2--0.8] keV data in black, the [0.8--2.4] kev data in dark grey
and the [2.4--10] keV data in light grey.
The model is characterised by a linear
increase and by an exponential decay, to which we add a Gaussian profile
reproducing the small flare at $\sim 19000$ s.
The peak position is fixed at 15000 s.
{\it Right panels}: we show the cross--correlations obtained from the
real 500 s binned light curves (dark grey) and from the curves 
in the left panel (light grey). 
CCF I: PN [0.8--2.4 keV] vs PN [0.2--0.8 keV].
CCF II: PN [2.4--10 keV] vs PN [0.2--0.8 keV].
In the inserts we show the peak regions.
}
\label{simul-decay} 
\end{figure}

Since XMM--Newton provides good temporal coverage for the
whole observing time, we performed the cross--correlations 
using the task CROSSCORR of the Xronos 5.19 package, 
based on a Fast Fourier algorithm which needs a continuous light curve,
without interruptions.
During the cross--correlation process, we filled the possible gaps
with the running mean value calculated over the 6 closest  bins.
We check the results with the Discrete cross--correlation technique
(DCC, Edelson \& Krolik 1988) to verify
the absence of distortions induced by the possible presence
of such small gaps (note, however, that the DCC does not provide 
an error estimate on the peak position).\\
We performed the cross--correlations on the whole light curves 
of November 14 and of December 1$^{\rm st}$ as well 
as on their main flares. Therefore, for the November 14 exposure, 
we excluded the first $\sim 25$ ks and the last $\sim 10$ ks,
while for the December 1$^{\rm st}$ we focused on the small feature,
lasting $\sim 14$ ks, occurring after about half observation.
Since the curves display several substructures,
as a check we performed cross--correlations also on the excluded  
subsections.\\
We compared the [0.2--0.8] keV with the [0.8--2.4] keV and the  
[2.4--10] keV light curves, using different temporal binning 
(50, 100, 200 and 500 s). In order to estimate the position of 
the CCF peaks, i.e. the delay amounts, we fitted them 
with a constant + a skewed Gaussian model 
(the $\sigma$ below and above the Gaussian peak are different).
This model, originally proposed and used by Brinkmann et al. (2003),
accounts for the possible asymmetries of the CCF 
and therefore it accurately constrains their maximum. 
For the November 14  cross--correlations we fitted the central $\pm 15$ ks 
part of the CCF, to investigate its overall shape. We also fitted only the 
$\pm 5$ ks central part to obtain a more accurate peak position. 
For the December 1$^{\rm st}$ observation, we fitted 
only the central $\pm 5$ ks.\\
We remark, however, that the peak position is not always a correct
delay estimator: for structured or asymmetrical light curves,
it does not represent adequately the real temporal behaviour.
In the first case, the possible delays in each variability event
will be mixed together and the resulting delay will be an average value,
obtained weighting each delay with its signal amplitude.
In the second case, the 
CCF asymmetry can be a more relevant
parameter, related to the slopes of the compared light curves (see below).\\
In Fig. \ref{crcor-1} we plot  the central peak of the cross--correlations 
performed on the main flares
of the November 14 and of the December 1$^{\rm st}$ light curves.
We also plot the Discrete cross--correlations
(light grey data) and the best--fit constant + skewed Gaussian 
models (solid black lines). 
In Table \ref{tab5} we report the best--fit peak positions
and the weighted average of the $\sigma_i$ parameters
for the  cross--correlations and for the Discrete cross--correlations.
Note, however, that the errors reported in Table \ref{tab5} are  underestimated
since they account only for the statistic uncertainties on the skewed Gaussian 
parameters, which are also affected by two kinds of windowing effects.
The first is related to the choice of the CCF
section to be fitted, while the second  is associated to the selection of the light 
curve intervals to be cross--correlated. Our simulations show that these effects
can introduce uncertainties on the peak positions as large as 200--300~s, which are 
probably a more realistic error estimation than what reported in Table~\ref{tab5}.
\begin{table*}
\begin{center}
\begin{tabular}{ccccc|cc}
\hline
Curves id.& \multicolumn{4}{c|}{Lag (sec)} & $<\sigma_1>$ & $<\sigma_2>$ \\
\hline
bin--time (s)  & 50 & 100 & 200 & 500 &($10^3$ s) & ($10^3$ s)  \\ 
\hline
\multicolumn{7}{c}{14 Nov 2002: whole curve. Central $\pm15$ ks.}\\
\hline
 CCF I  & $-890\pm40$ & $-750\pm60$ & $-830\pm70$ & $-760\pm120$ & $13.41\pm0.15$ & $12.76\pm0.17$  \\
 DCC I  & $-1030$ & $-830$  & $-740$ & $-930$ & $13.52$  & $12.72$  \\
 CCF II  & $-1860\pm60$ & $-1670\pm80$ & $-1730\pm110$ & $-1620\pm160$ & $15.04\pm0.15$ & $12.59\pm0.20$ \\
 DCC II  & $-2010$ & $-1760$ & $-1730$ & $-1920$ & $15.23$ & $12.36$ \\
\hline
\multicolumn{7}{c}{14 Nov 2002: whole curve. Central $\pm5$ ks.}\\
\hline
 CCF I  & $-140\pm90$ & $-60\pm110$ & $-50\pm180$ & $-20\pm250$ & $4.00\pm0.17$ & $2.81\pm0.20$  \\
 DCC I  & $-90$ & $-80$  & $-20$ & $-50$ & $4.04$  & $2.66$  \\
 CCF II   & $-180\pm130$ & $-180\pm180$ & $-120\pm260$  & $-90\pm360$ & $7.20\pm0.40$ & $3.02\pm0.27$ \\
 DCC II  & $-80$ & $-170$ & $-60$ & $-60$ & $7.40$ & $2.79$ \\
\hline
\multicolumn{7}{c}{14 Nov 2002: main flare. Central $\pm15$ ks.}\\
\hline
 CCF I & $-600\pm40$ & $-560\pm60$ & $-630\pm90$ & $-640\pm140$ & $9.04\pm0.06$ & $7.84\pm0.07$ \\
 DCC I  & $-770$ & $-700$ & $-620$ & $-790$ & $9.10$  & $8.01$ \\
 CCF II  & $-1390\pm60$ & $-1350\pm90$ & $-1340\pm130$ & $-1410\pm190$ & $10.46\pm0.07$ &  $8.24\pm0.09$ \\
 DCC II  & $-1580$ & $-1550$ & $-1460$ & $-1370$ & $10.39$ & $8.38$ \\
\hline
\multicolumn{7}{c}{14 Nov 2002: main flare. Central $\pm5$ ks}\\
\hline
 CCF I  & $-210\pm90$ & $-140\pm140$ & $-110\pm190$ & $-80\pm300$ & $3.97\pm0.18$ & $2.95\pm0.21$ \\
 DCC I  & $-220$ & $-160$ & $-90$ & $-150$ & $4.30$   & $3.14$ \\
 CCF II  & $-410\pm170$ & $-290\pm260$ & $-250\pm340$ & $-220\pm500$ & $8.31\pm0.56$ &  $4.01\pm0.43$ \\
 DCC II  & $-340$ & $-380$ & $-240$ & $-350$ & $8.58$ & $4.30$ \\
\hline
\hline
\multicolumn{7}{c}{1 Dec 2002: whole curve}\\
\hline 
CCF I  & $160\pm50$ & $220\pm60$ & $250\pm90$ & $230\pm120$ & $2.41\pm0.07$ & $4.04\pm0.06$ \\
DCC I &   $90$  & $160$ & $200$ & $120$ & $2.49$ & $4.03$ \\
CCF II  & $530\pm80$ & $570\pm100$ & $610\pm130$ & $740\pm220$ & $2.87\pm0.13$ & $6.04\pm0.18$\\
DCC II  &  $560$  & $570$ & $590$ & $400$ & $2.77$ & $6.08$\\
\hline
\multicolumn{7}{c}{1 Dec 2002: flare}\\
\hline
CCF I  & $430\pm50$ & $470\pm60$ & $490\pm80$ & $480\pm110$ & $3.29\pm0.10$ & $3.85\pm0.08$ \\
 DCC I  & $400$ & $460$ & $480$ & $450$ & $3.24$ & $3.83$ \\ 
CCF II  & $950\pm80$ & $950\pm100$ & $940\pm130$ & $860\pm180$ & $3.39\pm0.15$ & $4.34\pm0.10$\\
 DCC II  & $1040$ & $960$ & $920$ & $840$ &  $3.32$ & $4.29$\\
\hline
\hline
\end{tabular}
\caption{Best--fit parameters of the constant + skewed Gaussian model 
reproducing the cross--correlation (CCF) peaks. We cross--correlated
the whole light curves of November 14, of December 1$^{\rm st}$
and their main flares. We compared the [0.2--0.8] keV light curves 
to the  [0.8--2.4] keV (Id. I) and to the [2.4--10] keV light curves (Id. II).
We also performed a Discrete cross--correlation (DCC) on the same curves.
We reproduced the central $\pm 15$ ks and $\pm 5$ ks of the November 14 
CCF and the central $\pm 5$ ks of those of December 1$^{\rm st}$.
Negative lags mean that the variations in the hard X--ray band lead those 
in the soft X--ray band. The reported skewed Gaussian $\sigma_i$ 
 are the weighted averages of the four amplitudes.}
\label{tab5}
\end{center}
\end{table*} 
\begin{figure*}

\hbox to \textwidth
{\centerline{
\vbox {\psfig{figure=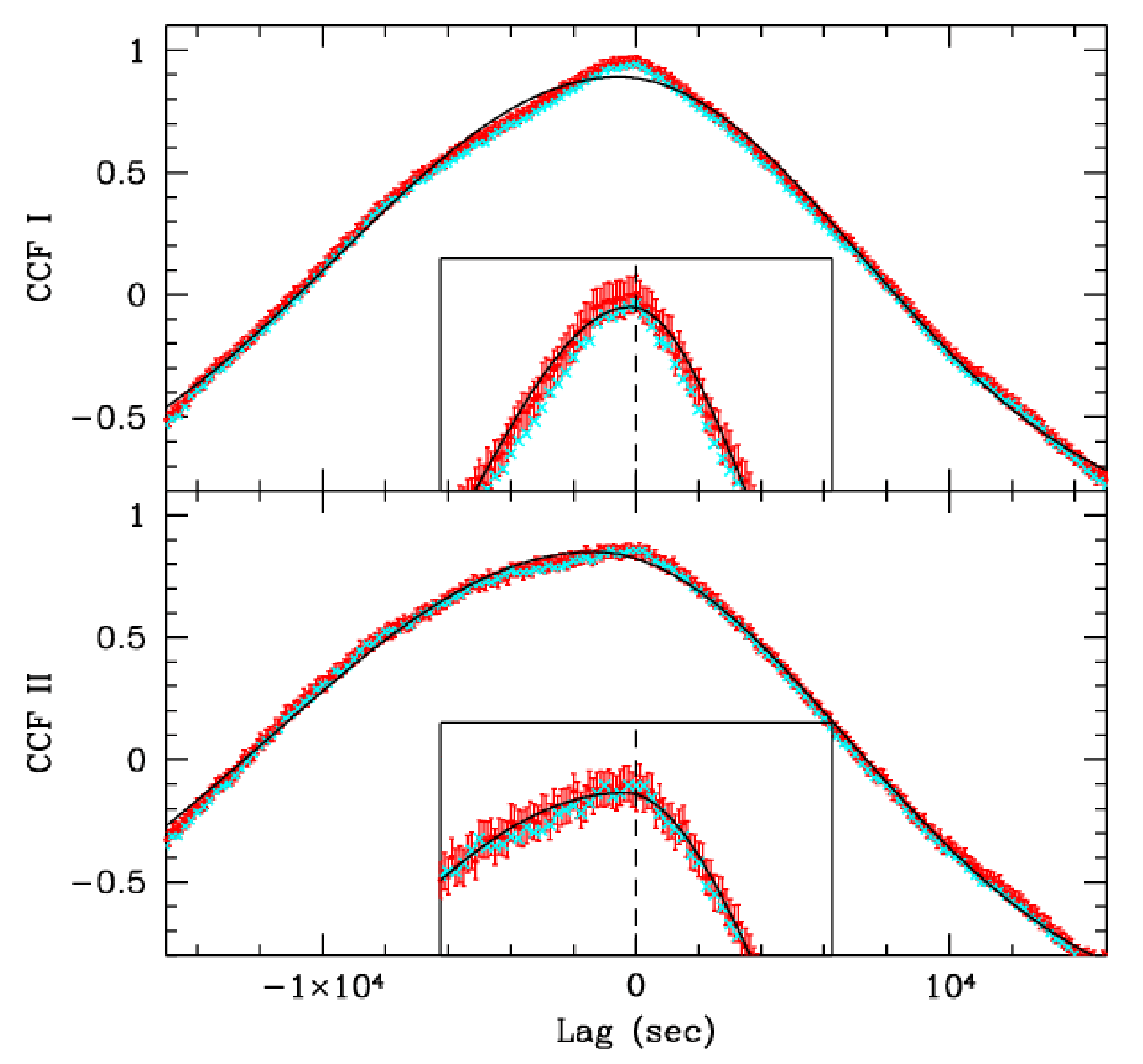,width=9.cm}}
\hskip -0.01cm
\vbox{\psfig{figure=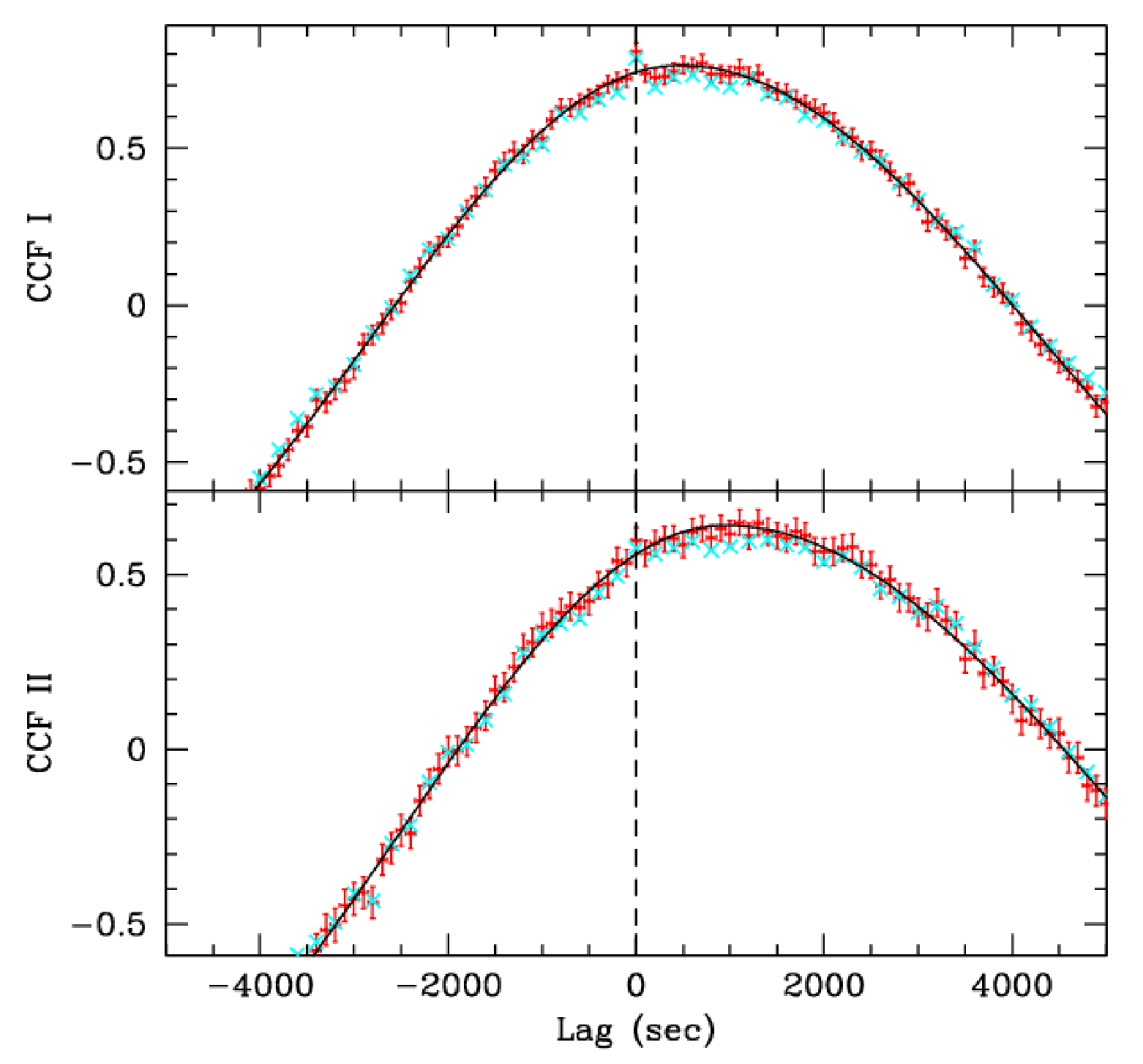,width=9.cm}}
}}
\caption{
The left panel shows the central $\pm 15$ ks of the 
cross--correlations performed on the flare of November 14. The insert 
shows only the central $\pm 5$ ks, with a better fit of the peaks.
The central $\pm 5$ ks of the cross--correlations performed on the
flare of  December 1$^{\rm st}$ is shown in the right panel.
The light curves were rebinned in 100 s bins.
CCF I: PN [0.8--2.4 keV] vs PN [0.2--0.8 keV].
CCF II: PN [2.4--10 keV] vs PN [0.2--0.8 keV].
We show in dark grey the cross--correlation data and in light grey
the Discrete cross correlation data. The solid black line is the best--fit
skewed Gaussian model (the insert data are fitted independently).
}
\label{crcor-1} 
\end{figure*}                  
We summarize the results of the cross--correlation analysis as follows:
\begin{itemize}
\item  
the results obtained from differently  binned light curves
are consistent with each other.
Furthermore, the Discrete cross--correlation results are fully consistent 
with those  of the cross--correlations.
The best--fit parameters relative to the two techniques, in fact,
are very similar
\item 
the lags obtained cross--correlating the whole curves are
similar (November 14) or smaller (December 1$^{\rm st}$) than those
obtained considering only the flares:
the delays are probably produced during the main flux variations.
This is confirmed by the absence of significant lags 
in the other sections of the light curves, as evidenced, e.g. 
by the lack of clear loop patterns in the hardness ratio
versus count rate plots corresponding to the minor flares of November 14 
(peaking at $\sim 10000$~s, $\sim 20000$~s and at $\sim 62000$~s),
by the cross--correlations performed on these intervals  
and by reproducing the curves with analytical models (see next Section)
\item 
November 14: the peak positions obtained for the $\pm 15$ ks
central part of the CCF are not consistent with zero delays. However, this is due
to the strong asymmetry of the CCFs that are not well fitted
even by a skewed Gaussian model. On this time interval the fit is
dominated by the wings of the CCF and the peak is not well fitted. A more accurate 
position of the peaks are obtained by fitting only the central $\pm 5$\,ks
of the CCFs. In this case the position of the peaks are consistent with 
$\sim$ zero delay (see the inserts
in the left panel of Fig.~\ref{crcor-1}). However, they are asymmetrical,
being broader toward negative lags. Thus, we have to explain
a CCF that has a zero lag delay, but an asymmetrical shape.
A possibility could be that this peculiar shape of the CCF is due to 
variability patterns present in both light curves, peaking 
simultaneously but with different rising and/or decaying time scales.
For instance, in the case of blazars we can imagine to have a flare
characterised by a linear increase, i.e. dominated by geometric effects,
followed by an exponential decay with different $\tau$ at different frequencies
(i.e. dominated by cooling effects).
To test this possibility, we generated simulated flare light curves, assuming 
different rising and decay time scales, but a simultaneous peak position 
for the flare in the two light curves. From their cross-correlation we obtained 
CCF that are very similar to the ones shown in Fig.~\ref{crcor-1}. We also added
a Gaussian feature to reproduce the small flare that is present in the real light 
curves at $\sim 19000$ s (see peak 3.2, next Section): this extra feature, however, 
has the same properties in all bands and does not introduce significant effects. 
Having shown that with such an analytical model (linear rising + exponential decay + 
a Gaussian feature) we can reproduce the observed CCF, we fitted it to our light curves.
With the best--fit parameters, we generated 500 s binned light curves, 
attributing to each point the uncertainty of the corresponding real one
(see Fig.~\ref{simul-decay}, left panels).
Then we cross--correlated these model--generated light curves and fitted 
the CCF peaks with a skewed Gaussian model as we did with the real light curves
(see Fig.~\ref{simul-decay}, right panels). The best--fit parameters are consistent 
with those reported in Table~\ref{tab5}. Similar results are obtained also using shorter 
temporal bins, so here we show only the case of the 500 s bins. Thus, with this simple
analytical model we can reproduce both the observed flare light curves and the resulting CCF.\\
Clearly, for this event the harder X--ray light curves have a steeper increase (i.e. a 
hardening of the spectrum) and a faster decay (i.e. a softening of the spectrum), leading 
on average those at softer energies. Therefore, even if the peaks are simultaneous, the 
different slopes of the flares will produce a sort of soft lag.
As a first indication of this lag, we will consider the difference between the halving 
times of the fitted exponential curves. In Table~\ref{simulag} we summarize our results.
\item 
December 1$^{\rm st}$: the cross--correlations
are  more symmetrical and their maxima are located at positive lags
(see the right picture of Fig.~\ref{crcor-1}).
Since this flare is quite smooth, the cross--correlation shapes
are probably originated by light curves peaking at
different times.  In this case, the peak positions of the best--fit skewed
Gaussian give a straightforward estimate of the delays.
During this flare, therefore,  the [0.2--0.8] keV variations lead those at
[0.8--2.4] keV and at [2.4--10] keV by $450\pm30$~s and by $950\pm60$~s,
respectively (these values are the weighted mean of those reported 
in Table \ref{tab5}). This behaviour can be produced, for instance, 
by an energy dependent particle acceleration:
lower energy particle are produced sooner (see the Discussion).
\item  
we confirm the results of the hardness ratio
and  time resolved spectral analysis: there are delays between 
flux variations at different energies.
During the November 14 observation, when the spectral evolution
was characterised by a clockwise loop pattern, the harder X--ray 
fluxes were decaying faster. During the small flare of December 
1$^{\rm st}$, instead, when we observed counterclockwise loop patterns, 
the mid and the hard X--ray flare peaks were delayed
by $450\pm30$ s and by $950\pm60$ s, respectively.
\item 
comparing light curves of different energy range, the delays are larger 
for a larger difference between the energy ranges considered:
the temporal lags between the [0.2--0.8] keV and 
the [2.4--10] keV light curves are larger than those 
between the [0.2--0.8] keV and the [0.8--2.4] keV curves.
\end{itemize}
\begin{table}
\begin{center}
\begin{tabular}{cccc}
\hline
Energy band & e-folding time & Halving time & Lag \\
 (keV)      & ($10^4$ s) & ($10^4$ s) & ($10^4$ s) \\
\hline 
$[0.2-0.8]$ & $25.2\pm0.9$ & $17.5\pm0.6$ &       \\ 
$[0.8-2.4]$ & $16.9\pm0.8$ & $11.7\pm0.6$ & $-5.8\pm0.8$ \\
$[2.4-10]$  & $8.3\pm1.1$ & $5.8\pm0.8$  & $-11.6\pm1.0$ \\
\hline
\end{tabular}
\caption{
Best-fit e-folding time of the exponential model reproducing
the flare decay of November 14 in the three energy bands. We also report the 
respective halving times and their differences between the soft and the
two harder energy bands.
}
\label{simulag}
\end{center}
\end{table}
In the next Section we will check these 
results by fitting the light curves with analytical models.
\subsection{Modelling the light curves}
\begin{figure}
\centerline{
\vbox{
\psfig{figure=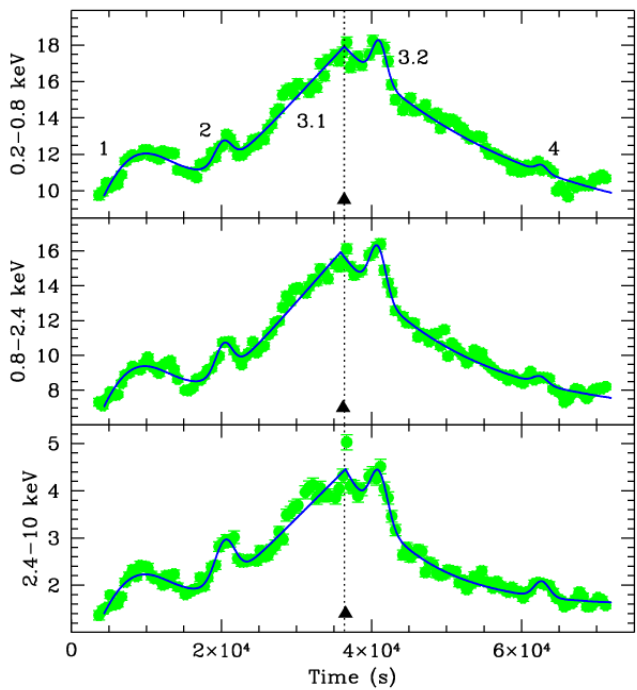,width=9cm}
}}
\caption{November 14, 2002: PN [0.2--0.8] keV, [0.8--2.4] keV and [2.4--10] keV
500 s rebinned light curves of Mkn 421 (the y--axis unit is count/s).
We plot the best--fit model as a solid black line:
we used a linear increase+exponential decay curve summed 
to 4 Gaussian profiles.
The black filled triangles represent the peak position of flare 3.1
in the three bands: they are nearly simultaneous.
}
\label{5gauss} 
\end{figure}

\begin{figure}
\centerline{
\vbox{
\psfig{figure=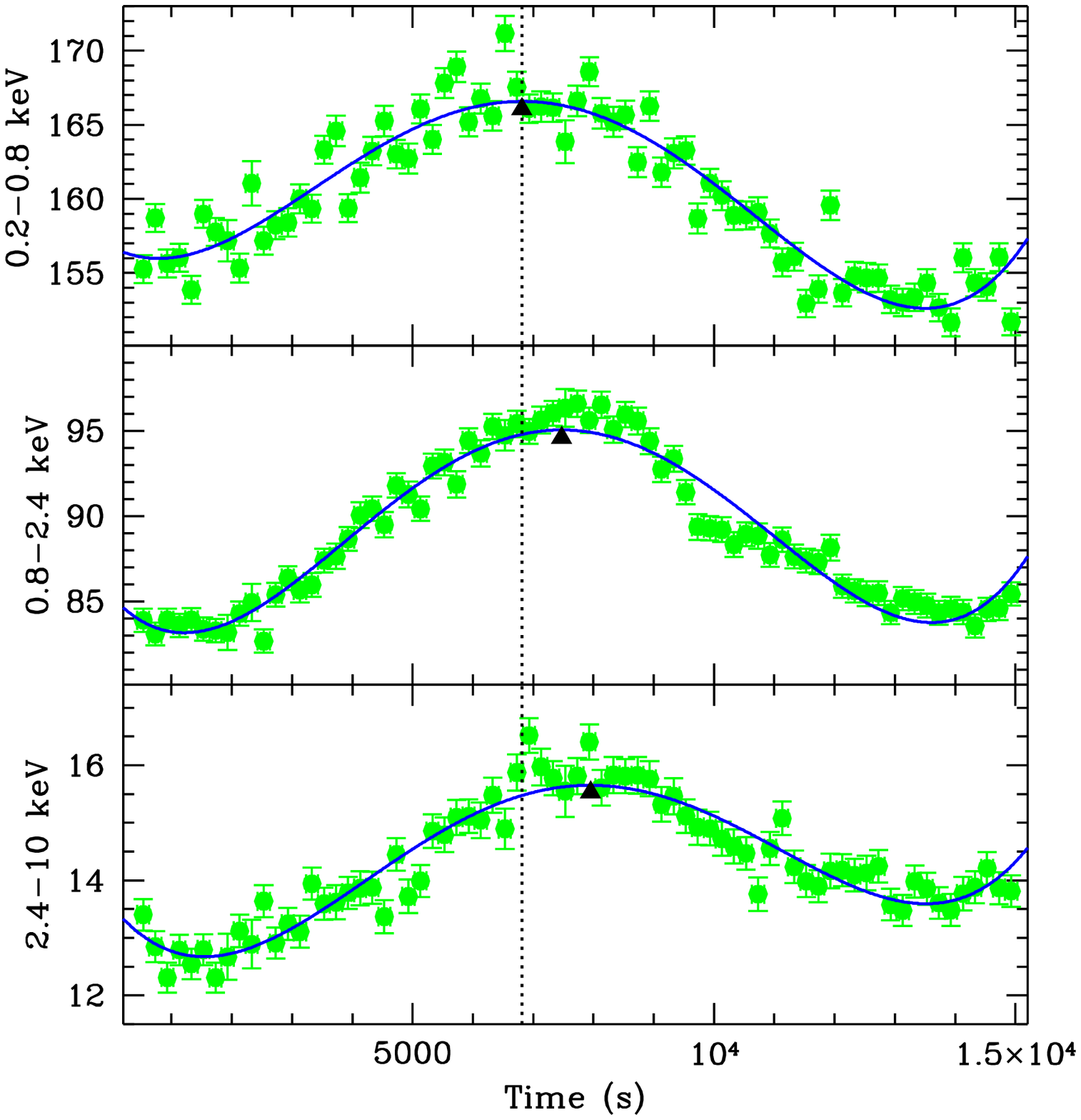,width=9cm}
}}
\caption{December 1$^{\rm st}$, 2002: PN [0.2--0.8] keV, [0.8--2.4] keV and 
[2.4--10] keV 200 s rebinned light curves of Mkn 421. We plot the 
small flare detected at about half observation. The solid line represents
 the best--fit 4$^{\rm th}$ degree polynomial model. The dotted vertical line
represents the [0.2--0.8] keV  peak position and the black filled triangles
indicate the peak position of the three curves. It is clear
 that the  high energy curve peaks are delayed.}
\label{polyn} 
\end{figure}

We rebinned the November 14 light curve and the  December 1$^{\rm st}$
flare using 500 s and 200 s bins, respectively.
Since the November 14 curve was very structured, we fitted it
with the linear increase+exponential decay model  described in the previous 
Section + 4 Gaussian profiles (see Fig. \ref{5gauss}).
The asymmetrical curve and one Gaussian were aimed at reproducing 
the large central flare (henceforth peak 3): 
the first (peak 3.1 in Fig. \ref{5gauss})
representing the main, average variation
and the second one describing the clear bump at $\sim 41000$ s (peak 3.2).
The other Gaussian were used to model  the small features
at $\sim 10000$ s (peak 1), $\sim 20000$~s (peak 2)
and  $\sim 62000$ s (peak 4) from  the beginning of the observation.
We were able to reproduce the light curves 
leaving all the parameters free to vary in the best fit procedure.\\
The December 1$^{\rm st}$ flare, instead,
was quite smooth and we fitted it
with a 4$^{\rm th}$ degree polynomial peaking at $\sim 11000$ s from 
the beginning of the temporal window (see Fig.\ref{polyn}).
We chose this profile because it well reproduce the light curve
asymmetries. However, to estimate the uncertainties on the peak positions,
we fitted the flare also with a constant plus a Gaussian model.
The results obtained with the two models are  very similar.
In Table \ref{flarefit-lags} we report all the  best--fit parameters.\\
\begin{table*}
\begin{center}
\begin{tabular}{cccccc}
\hline
\multicolumn{6}{c}{14 November}\\
\hline
Feature & \multicolumn{3}{c}{Time} & \multicolumn{2}{c}{Lag} \\
        & [0.2--0.8] keV & [0.8--2.4] keV & [2.4--10] keV & $t_2-t_1$ & $t_3-t_1$ \\
        &  (s) & (s) & (s) & (s) & (s) \\
\hline
Peak 1  & $6990^{+300}_{-330}$ & $6810^{+240}_{-270}$ & $6980^{+340}_{-400}$  & $-180^{+380}_{-430}$ &  $-10^{+450}_{-520}$  \\
Peak 2  & $20220^{+250}_{-260}$ & $20330^{+160}_{-170}$ & $20470^{+160}_{-180}$ & $+110^{+300}_{-310}$ & $+250^{+300}_{-320}$ \\
Peak 3.1 & $36360^{+330}_{-430}$ & $36220^{+240}_{-640}$ & $36550^{+220}_{-410}$ & $-140^{+410}_{-770}$ & $+190^{+400}_{-590}$ \\
$\sigma_1$ & $25170^{+980}_{-910}$ & $16260^{+880}_{-810}$ & $9560^{+800}_{-680}$ \\
Peak 3.2 & $41040\pm220$ & $40860\pm130$ & $41000\pm200$ & $-140\pm260$ & $-40\pm300$ \\    
Peak 4   & $62680^{+670}_{-560}$ & $62860^{+530}_{-480}$ & $62670\pm290$ & $+180^{+850}_{-740}$ & $-10^{+730}_{-630}$ \\
\hline
\hline
\multicolumn{6}{c}{1 December}\\
\hline
Flare start & $790$  & $1210$  & $1550$ & $+420$ & $+760$  \\
Flare peak  & $6810$ & $7470$ & $7950$ & $+660$ & $+1140$  \\
Peak (Gauss.) & $6760\pm100$ & $7430\pm80$ & $8010\pm160$ & $+670\pm130$ & $+1250\pm180$\\
Flare end  & $13490$ & $13610$ & $13510$ & $+120$  & $+10$      \\
\hline
\end{tabular}
\caption{Maxima and minima of the best--fit models of the 
November 14 and of the December 1$^{\rm st}$ flares 
in three different energy bands. The November 14 and the December 1$^{\rm st}$ 
light curves are binned in 500 s and 200 s intervals, respectively.
For the November 14 observation we report also the best-fit e-folding 
time of the exponential decay model.
For the December observation we give the start and stop time of the flare
and the flare peak (using two different models, see text).
In columns 5 and 6 we report the delays between the features
of the medium [0.8--2.4] keV (Col. 5)
and the hard [2.4--10] keV (Col. 6) light curves with respect 
to those of the soft [0.2--0.8] keV light curve.}
\label{flarefit-lags}
\end{center}
\end{table*}
During the November 14 observation the source behaved in a complex way:
\begin{itemize}
\item
both peak 1 (at $\sim 10000 \, $s) and peak 2 (at $\sim 20000 \,$s)
occur almost simultaneously in the three bands; all delays are consistent
with zero.
\item 
even the two subcomponents forming the third, large flare 
do not show significant peak shifts. 
We remark, however, the large differences in the 
decay slopes of the peak 3.1. 
Although the flare peaks are almost simultaneous, 
the flux increase is larger in the harder X--ray bands
and the following decay is much faster.
As shown in the previous section, this produces the distortions 
observed in the cross--correlations.
\item 
again the small peak 4 occurs almost simultaneously in the three
bands, although with larger uncertainties due to the fact that this event is 
well pronounced in the harder energy band but not in the other two.
\end{itemize}
The absence of delays at the peaks 1, 2 \& 4 is confirmed by the 
lack of loop patterns in the corresponding 
hardness ratio vs count rate plots.
We conclude therefore that the clockwise loop patterns
evidenced in Section 4.1 and Section 4.2 are connected 
with the presence of soft lags, mainly caused by 
the different slopes of the peak 3.1.
The smaller substructures are not 
characterised by the presence of significant delays.
As expected (see previous section), we find that the slope difference
between the [0.2--0.8] keV and the [2.4--10] keV light curves is  larger 
than that between the [0.2--0.8] keV and the [0.8--2.4] keV light curves.\\
The situation is different for the isolated  flare of December 1$^{\rm st}$:
the [0.2--0.8] keV leads the mid and the hard curves both at the beginning
($\sim 420$ s and $\sim 760$ s, respectively)
and at the peak of the flare ($\sim 660$ s and $\sim 1140$ s),
as confirmed also by the constant + Gaussian model.
The delay of the [2.4--10] keV variation is significantly larger 
than that of the [0.8--2.4] keV curve and 
they are consistent with those obtained through the cross--correlations.
The fade of the flare, instead, seems to stop almost simultaneously in the three bands. \\
The light curves are very structured and our models do
not exactly follow the small substructures that are present.
However, the use of more complex  models is beyond our goal, that is to determine 
the existence and the amount of delays between the main variability features 
in different energy bands.
The lags reported in Table \ref{flarefit-lags} are therefore average 
values mixing the contributions of the light curve substructures,
in line with the results obtained from the cross-correlation analysis.
\section{Discussion}
We observed X--ray  spectral evolution during two complete flares of Mkn 421
through a hardness ratio and a time resolved spectral analysis.
This was clearly shown by the presence of hysteretic patterns in the hardness 
ratio vs count rate plots and in the spectral index vs flux plots. 
Such characteristic patterns  are usually explained as the signatures
of temporal delays between different energy light curves 
(see e.g. Takahashi et al. 1996).
Then, we used two techniques to check the reality and 
to estimate the amount of such possible lags;
a) we performed  a cross--correlation analysis
and b) we reproduced the light curves with analytical models to compare
the positions of the maxima.
We confirmed the presence of the temporal lags. More precisely, we found soft lags 
in the observation of November 14, produced by different variability rates during 
a single, even if structured flare (peak 3.1 in Fig. \ref{5gauss}), which cannot 
be further split. In the observation of December 1$^{\rm st}$
we observed the opposite behaviour: 
a small, smooth flare is characterised by large hard lags.
We found also that the delays between the [0.2--0.8] keV 
and the [2.4--10] keV bands are larger than those between the [0.2--0.8] keV 
and the [0.8--2.4] keV bands. They must be produced  
by energy dependent mechanisms like, for instance,
the particle cooling and acceleration.\\
Following the treatment of Zhang et al. (2002a), 
we can express the cooling timescale $t_{cool}$ and the acceleration
timescale $t_{acc}$ in the observer frame 
as a function of the photon energy $E$ (in keV) as:
\begin{equation}
t_{cool}(E) = 3.04 \times 10^3 (1+z)^{1/2} B^{-3/2} \delta^{-1/2} E^{-1/2} s
\label{tcool}
\end{equation}
\begin{equation}
t_{acc}(E) = 9.65 \times 10^{-2} (1+z)^{3/2} \xi B^{-3/2} \delta^{-3/2} E^{1/2} s
\label{tacc}
\end{equation}
where $z$ is the redshift of the source, $B$ is the magnetic field in Gauss,
$\delta$ is the Doppler factor of the emitting region and $\xi$ is a
parameter indicating the acceleration rate of electrons
(see Zhang et al., 2002a). As evidenced by  equations 
\ref{tcool} and \ref{tacc}, the cooling and the acceleration mechanisms
behave oppositely with respect to the photon energy $E$:
higher energy particles cool faster and accelerate slower.
Another important timescale which could be involved in the 
production of  the  delays
is the light crossing time  of the emitting region $t_{esc}$.
In fact,  Ghisellini, Celotti \& Costamante (2002)
suggested that the synchrotron peak of HBL objects
(and therefore of Mkn 421) is produced by particles with 
cooling time $t_{cool} = t_{inj} \sim t_{esc}$,
where $t_{inj}$ is the particle injection/acceleration timescale. 
In an internal shock scenario $t_{inj}$ is very similar to $t_{esc}$.\\
A different balancing of these characteristic timescales,
$t_{cool}$, $t_{acc}$ and $t_{esc}$
can account for the observed temporal lags.
\begin{itemize}
\item 
November 14: the November 14  light curve is very structured,
showing  several small features.
However, only the large flare 3 is characterised by 
clear lags, mainly caused by the different slopes of the peak 3.1.
Since the source is displaying a soft spectrum above 0.6 keV,
with spectral index  $\alpha \sim 1.13$, we are very close to the synchrotron peak, 
which could be even located inside our softer  energy  band ([0.2--0.8] keV).
This implies that $t_{cool} \sim t_{esc} >> t_{acc}$
(since $t_{acc} = t_{cool}$ at  the highest observed synchrotron energy $E_{max}$).\\
In this case, we assume a particle acceleration,
that produces a spectral hardening and leads to simultaneous peaks,
followed by a decay dominated by cooling effects.
Since the highest energy particles suffer the quickest cooling,
we will observe soft lags and clockwise loop patterns in the 
spectral index vs flux plots. The  soft lags and their 
frequency dependence observed during this flare can be attributed
to the  frequency dependence of $t_{cool}$.
\item 
December 1$^{\rm st}$: the small flare after
about half observation shows large hard lags.
In this case the source spectrum is softer 
($\alpha\sim 1.5$) than in  November 14.
We are therefore closer to $E_{max}$, where
$t_{cool} \sim t_{acc}$. In this case we can assume
$t_{esc} >> t_{cool} \sim t_{acc}$: 
the information about the occurrence of a flare propagates from 
lower to higher energies, as particles are gradually accelerated,
while the decay of the flare could be dominated by the particle
escape effects, which can be assumed achromatic.
Then we will observe hard lags and counterclockwise
loop patterns, produced by real delays at the peak of the flares.
In this case, the observed hard lag will be generated
by the frequency dependence of $t_{acc}$.\\ 
This scenario is supported by the shape of the loop patterns 
shown in Fig. \ref{clock} (right panels):
as the flux begins to increase, the spectrum softens.
This can be explained as an effect of the progressive acceleration:
the spectrum initially steepens because  electrons 
cannot be accelerated to higher energies, yet.
\end{itemize}
The presence of  soft and  hard lags   can therefore be explained
in the framework of different cooling or acceleration timescales.
A similar conclusion was reached also by other authors, 
by solving the particle and photons continuity equations
(see e.g. Kirk, Rieger \& Mastichiadis 1998).\\
The detection of lags  can shed some light
on the acceleration as well as on the cooling mechanisms and
provide a powerful tool to constrain the physical parameters of the source.
In fact, if the  soft lags $\tau_{soft}$ of November 14
are produced by cooling effects, 
($\tau_{soft} = t_{cool}(E_s) - t_{cool}(E_h)$) and the hard lags 
of December 1$^{\rm st}$ are produced by acceleration effects 
($\tau_{hard} = t_{acc}(E_h)-t_{acc}(E_s)$), we can estimate the 
physical properties of the emitting region through the equations
\begin{equation}
B~ \delta^{1/3} = 209.91 \Big(\frac{1+z}{E_s}\Big)^{1/3} \Big[\frac{1-(E_s/E_h)^{1/2}}{\tau_{soft}}\Big]^{2/3} G
\label{eq-cool}
\end{equation}
\begin{equation}
B~ \delta~ \xi^{-2/3} = 0.21 (1+z) E^{1/3}_h \Big[\frac{1-(E_s/E_h)^{1/2}}{\tau_{hard}}\Big]^{2/3} G
\label{eq-acc}
\end{equation}
where $E_s$ and $E_h$ are the mean energies in the corresponding energy bands, 
taking into account the power--law shape of the spectrum 
(Zhang et al., 2002a).\\
Since the amount of the lags changes when comparing
different couples of light curves,
we have  the opportunity to check the reality of this scenario.
If our assumptions are correct,
assuming that a flare is produced by a single electron population, 
using the  lags between different  couples of light curves 
in the equations \ref{eq-cool}
and \ref{eq-acc}, we should obtain the same emitting region characteristics.
For the flare of November 14, we used the delays obtained from the halving 
times differences of the simulated light curves,
while for that of December 1$^{\rm st}$ we used the difference 
between the cross--correlation peak positions.
In Table \ref{tab6} we report the assumed parameters and the results.
For the December 1$^{\rm st}$ lags, we do not consider 
the uncertainties  shown in Table \ref{tab5} since they are probably
underestimated: we will assume more conservative error values of 200~s.\\
\begin{table*}
\begin{center}
\begin{tabular}{ccccccc}
\hline
$E_1$ & $E_2$ & $E_3$ & $\tau_{12}$ & $\tau_{13}$ & $B_{12}$ &
 $B_{13}$ \\
(keV) & (keV) & (keV) & (s) & (s) & (G) & (G) \\
\hline
\multicolumn{7}{c}{November 14 soft lag}\\
\hline
0.42 & 1.44 & 5.19 & 5800 & 11600  & $0.24\pm0.02~\delta_{10}^{-1/3}$ & $0.21\pm0.01~\delta_{10}^{-1/3}$ \\
\hline
\multicolumn{7}{c}{December 1 hard lag}\\
\hline
0.40 & 1.38 & 4.87 & $450\pm200^*$ & $950\pm200^*$ & $0.53\pm0.12 ~\delta_{10}^{-1} \xi_5^{2/3}$ & $0.65\pm0.08 ~\delta_{10}^{-1} \xi_5^{2/3}$\\
\hline
\end{tabular}
\caption{To evaluate the mean energies of the three analysed bands, 
we used the spectral indexes of the best--fit power--law models
in the [0.6--10] keV ranges 
($\alpha = 1.15\pm0.01$ and $\alpha = 1.535\pm0.003$). 
We do not report the errors on the mean energies which are of the order 
of $10^{-4}$ keV.
The notations $\delta_{10}$ and $\xi_5$ mean ($\delta/10$) and
$\xi/10^5$, respectively. $^*$: since the uncertainties reported
 in Table \ref{tab5} are probably underestimated, we assumed more 
conservative errors of 200~s.
}
\label{tab6}
\end{center}
\end{table*}
The data reported in Table \ref{tab6} are consistent 
with the proposed scenario:
the observed soft lags are likely to be produced by the particle cooling 
and the hard lags by a progressive acceleration.
The difference between the magnetic fields obtained from the November 14
and for the December 1$^{\rm st}$ data, 
probably reflects our poor knowledge on the details 
of the real particle acceleration mechanism working in blazars.\\
It is interesting to point out that the magnetic field values 
reported in Table \ref{tab6} ($B\sim 0.2-0.65$ G)
are higher than those obtained
modelling the multiwavelength SEDs of the source
with SSC models. These models, in fact,  require weak magnetic fields  
to reproduce the observed TeV emission (e.g. Ghisellini, Celotti \& Costamante 2002).
This inconsistency could be caused by the techniques employed to 
estimate the lags, which provide only  lower limits of the ``real'' delays
when applied to light curves displaying substructures
with different behaviours, or by the poor knowledge of the acceleration
parameter $\xi$ (which we arbitrarily assumed to be $10^5$
in the case of the smooth flare of December 1$^{\rm st}$).

\section{Conclusion}

We  presented the spectral and temporal analysis of 3 XMM--Newton
 observations of Mkn 421. We resume here the main results:
\begin{enumerate}
\item The X--ray spectra of Mkn 421 are soft and steepen
toward higher energies: the November 4 spectra are best fitted 
by a softening  parabolic model, while the November 14 
and the December 1$^{\rm st}$ data are best fitted
by convex broken power--laws.
We are probably observing synchrotron emission from a range
above the low energy peak of the SED,
which, however, should be located very close to our lower limit
(e.g. in November 14, $\alpha_1 = 1.13$).
\item The hardness ratio analysis of two complete, 
different flares occurring in November 14 and in December 1$^{\rm st}$
shows the presence of strong spectral evolution. 
Besides presenting a clear harder--when--stronger 
correlation,  the hardness ratio vs count rate plots display 
characteristic loop patterns, which are the signature of temporal delays
between flux variations in different energy bands.
During the November 14 flare, 
the loop pattern rotates clockwise, suggesting the presence
of soft lags (see e.g. Takahashi et al. 1996), 
while during the December 1$^{\rm st}$ flare, the loop pattern rotates 
counterclockwise (hard lags, see e.g. Fossati et al. 2000b). 
These results were also confirmed by Brinkmann et al. (2003) using
high quality XMM--Newton data.
\item We confirmed the results of the hardness ratio analysis performing
a time resolved spectral analysis. We observed again 
the loop patterns rotating clockwise and counterclockwise
in  November 14 and in the flare
of December 1$^{\rm st}$, respectively.
\item We verified the presence of the delays 
performing a cross--correlation analysis. 
We found that the lags are mainly produced by the  complete flares
of November 14 and December 1$^{\rm st}$,
while the rest of the light curves do not show delays.
In the first case, the flare peaks are simultaneous but are
characterised by different slopes, producing, on the average, soft lags.
In the second case, the flare peaks display significant hard lags.
The clockwise loop patterns are then associated with the presence 
of soft lags, while the counterclockwise loop patterns 
are associated with hard lags. We also found that the delays  increase
with the energy difference between the compared light curves.
\item
We fitted the November 14 and the December 1$^{\rm st}$ flares
at different energies with analytical light curve models, to split them 
in their subcomponents. We estimated the delays for each
component obtaining agreement with  the results of the
cross--correlation and of the hardness ratio analysis.
The main flare of  November 14 does not  display  peak delays,
but it is characterised by different slopes, producing,
on the average, soft lags. The other components of this
light curve do not show significant delays.
The December 1$^{\rm st}$ flare is characterised by hard lags.\\
The complex behaviours of the subcomponents can be explained 
as produced by different emitting regions.
This is naturally accounted by the internal shock model 
proposed by Ghisellini (1999) and by Spada et al. (2001).
\item We presented a scenario to explain the presence of 
soft or hard lags as a consequence of different cooling and acceleration 
timescales. The results of the data analysis
are quite consistent with this picture,
suggesting that the frequency dependence of the synchrotron cooling 
is probably responsible for the November 14 soft lags. 
Also the hard lags in the December 1$^{\rm st}$ flare are
roughly compatible with  the assumed acceleration mechanism.
\end{enumerate}
We demonstrated that the hardness ratio and the 
temporally resolved spectral analysis are very powerful tools
to establish the presence of temporal lags between light curves at 
different energies. With the cross--correlation technique
we were able to estimate the amount of the delays. It is however
important to point out that this technique must be used cautiously.
While it is very reliable when applied to single smooth 
and symmetrical flares,
it can produce mixed results when applied to the complex 
and structured light curves of blazars as a whole.
A careful check of the behaviour of the single components
must be performed
before using the cross--correlation results to test the blazar models.
\begin{acknowledgements}
We thank the referee, W. Brinkmann, for comments that helped us to improve 
an earlier version of the paper, in particular for a better understanding 
of the cross correlation analysis results.
This research was financially supported by the Italian Space Agency and 
by the Italian Ministry for University and Research. 
\end{acknowledgements}


\begin{thebibliography}{ }
\small
\bibitem[Blazejowski et al., 2000]{bla} Blazejowski, M., Sikora, M., Moderski, R. \& Madejski, G.M. 2000, ApJ, 545, 107
\bibitem[B\"ottcher et al., 2003]{bot} B\"ottcher, M., Marscher, A.P., Ravasio, M. et al. 2003, ApJ, in press.
\bibitem[Brinkmann et al., 2001]{bri} Brinkmann, W., Sembay, S., Griffiths, R.G.  et al. 2001, A\&A, 365, L162
\bibitem[Brinkmann et al., 2003]{bri2} Brinkmann, W, Papadakis, E., den Herder, J.W.A. \& Haberl, F. 2003, A\&A, 402, 929
\bibitem[Chiaberge \& Ghisellini, 1999]{chi} Chiaberge, M. \& Ghisellini, G. 1999, MNRAS, 306, 551
\bibitem[Dermer \& Schlickeiser, 1993]{der} Dermer, C.D. \& Schlickeiser, R. 1993, ApJ, 416, 458
\bibitem[Edelson \& Krolik, 1988]{ede} Edelson, R. \& Krolik, J. 1988, ApJ, 333, 646 
\bibitem[Fossati et al., 2000a]{fos} Fossati, G., Celotti, A., Chiaberge, M. et al. 2000, ApJ, 541, 153
\bibitem[Fossati et al., 2000b]{fos} Fossati, G., Celotti, A., Chiaberge, M. et al. 2000, ApJ, 541, 166
\bibitem[Fossati, 2001]{fos} Fossati, G. 2001, in X--ray Astronomy 2000, 
Palermo, September 2000, ASP Conf. Ser., ed. R. Giacconi, L. Stella \& S. Serio
(San Francisco:ASP)
\bibitem[Gaidos et al., 1996]{gai} Gaidos, J.A., Akerlof, C.W., Biller, S.D. et al. 1996, Nature, 383, 319
\bibitem[Ghisellini \& Madau, 1996]{ghi} Ghisellini, G. \& Madau, P. 1996, MNRAS, 280, 67 
\bibitem[Ghisellini, 1999]{ghi3} Ghisellini, G. 1999, 4$^{th}$ ASCA Symp., Astronomische Nachrichten, H. Inoue, T. Ohashi \& T. Takahash eds., 320, 232
\bibitem[Ghisellini, Celotti \& Costamante, 2002]{ghi2} Ghisellini, G., Celotti, A. \& Costamante, L. 2002, A\&A, 386, 833
\bibitem[Giommi et al., 2000]{gio} Giommi, P., Padovani, P. \& Perlman, E. 2000, MNRAS, 317, 743 
\bibitem[Guainazzi et al., 1999]{gua} Guainazzi, M., Vacanti, G., Malizia, A. et al. 1999, A\&A, 342, 124
\bibitem[Kataoka et al., 2000]{kat} Kataoka, J., Takahashi, T., Makino, F. et al. 2000, ApJ, 528, 243
\bibitem[Kino et al., 2002]{kin} Kino, M., Takahara, F. \& Kusunose, M. 2002, ApJ, 564, 97
\bibitem[Kirk, Rieger \& Mastichiadis, 1998]{kir} Kirk, J.G., Rieger, F.M. \& Mastichiadis, A. 1998, A\&A, 333, 452
\bibitem[Krawczynski H. et al.,2001]{kra} Krawczynski, H., Sambruna, R., Kohnle, A. et al. 2001, ApJ, 559, 187 
\bibitem[Lockman \& Savage, 1995]{loc} Lockman, F.J. \& Savage, B.D. 1995, ApJS, 97, 1
\bibitem[Malizia et al., 2000]{mal} Malizia, A., Capalbi, M., Fiore, F. et al.  2000, MNRAS, 312, 123
\bibitem[Maraschi, Ghisellini \& Celotti, 1992]{mar2} Maraschi, L., Ghisellini, G. \& Celotti, A. 1992, ApJ, 397, L5
\bibitem[Maraschi et al., 1999]{mar} Maraschi, L., Fossati, G., Tavecchio, F. et al. 1999, ApJ, 526, L81
\bibitem[Massaro et al., 2003a]{mas} Massaro, E., Giommi, P., Tagliaferri, G. et al. 2003a, A\&A, 399, 33
\bibitem[Massaro et al., 2003b]{mas2} Massaro, E., Perri, M., Giommi, P., Nesci, R. 2003b, submitted to A\&A
\bibitem[Mastichiadis \& Kirk, 1997]{mast} Mastichiadis, A. \& Kirk, J.G. 1997, A\&A, 320, 19
\bibitem[Pian et al., 1998]{pian} Pian, E., Vacanti, G., Tagliaferri, G. et al. 1998, ApJ, 492, L17
\bibitem[Punch et al., 1992]{pun} Punch, M., Akerlof, C.W., Cawley, M.F. et al. 1992, Nature, 358, 477
\bibitem[Ravasio et al. 2002]{rav} Ravasio, M., Tagliaferri, G., Ghisellini, G. et al. 2002, A\&A, 383, 763
\bibitem[Sembay et al., 2002]{sem} Sembay, S., Edelson, R., Markowitz, A. et al. 2002, ApJ, 574, 634
\bibitem[Sikora, Begelmann \& Rees, 1994]{sik} Sikora, M., Begelmann, M.C. \& Rees, M.J. 1994, ApJ, 421, 153
\bibitem[Spada et al., 2001]{spa} Spada, M., Ghisellini, G., Lazzati, D. \& Celotti, A. 2001, MNRAS, 325, 1559
\bibitem[Takahashi et al., 1996]{tak} Takahashi, T., Tashiro, M., Madejski, G. et al. 1996, ApJ, 470, L89
\bibitem [Takahashi et al., 2000]{tak2} Takahashi, T., Kataoka, J., Madejski, G. et al. 2000,  ApJ,  542, L105 
\bibitem[Urry \& Padovani, 1995]{urry} Urry, M.C. \& Padovani, P. 1995, PASP, 107, 803
\bibitem[Zhang et al., 2002a]{zha} Zhang, Y.H., Treves, A., Celotti, A. et al. 2002a, ApJ, 572, 762
\bibitem[Zhang, 2002b]{zha} Zhang, Y.H. 2002b, MNRAS, 337, 609
\end{thebibliography}
\end{document}